\documentclass[pra,superscriptaddress,twocolumn,floatfig,showpacs,floatfix]{revtex4-1}

\usepackage{graphicx,color}
\usepackage{amsmath,amssymb,bm}
\usepackage{braket}	

\usepackage{mathtools}
\usepackage{tikz}
\usetikzlibrary{calc,arrows.meta,positioning}
\usetikzlibrary{arrows,positioning}
\usetikzlibrary{shapes,snakes}
\usepackage{float}
\usepackage[plainpages=false,pdfpagelabels,colorlinks=true,
linkcolor=blue,urlcolor=blue,citecolor=blue,pdftitle={Thermal and optical conductivity in the Holstein model at half filling and at finite temperature in the Luttinger-liquid and charge-density-wave regime},
pdfauthor={},pdfdisplaydoctitle=true,pdfduplex=DuplexFlipLongEdge]{hyperref}

\definecolor{darkred}{rgb}{0.90,0,0}
\definecolor{darkgreen}{rgb}{0,0.60,.2}
\definecolor{darkblue}{rgb}{0,0,1}
\definecolor{grey}{cmyk}{0,0,0,0.25}
\definecolor{orange}{cmyk}{0,0.6,1,0}

\usepackage{physics}
\usepackage{mathrsfs}
\newcommand{\beginsupplement}{%
        \setcounter{table}{0}
        \renewcommand{\thetable}{S\arabic{table}}%
        \setcounter{figure}{0}
        \renewcommand{\thefigure}{S\arabic{figure}}%
                \setcounter{equation}{0}
        \renewcommand{\theequation}{S.\arabic{equation}}
     }
\begin{document}
\title{
Thermal and optical conductivity in the Holstein model at half filling and at finite temperature in the Luttinger-liquid and charge-density-wave regime
}

\author{David Jansen}
\affiliation{Institut f\"ur Theoretische Physik, Georg-August-Universit\"at G\"ottingen, 37077 G\"ottingen, Germany}
\affiliation{ICFO-Institut  de  Ciencies  Fotoniques,  The  Barcelona  Institute  of  Science  and  Technology, 08860 Castelldefels (Barcelona), Spain}
\author{Fabian Heidrich-Meisner}
\affiliation{Institut f\"ur Theoretische Physik, Georg-August-Universit\"at G\"ottingen, 37077 G\"ottingen, Germany}

\begin{abstract}
Electron-phonon interactions play a key role in many branches of solid-state physics. Here, our focus is on the transport properties of one-dimensional systems, and we apply efficient real-time matrix-product state methods to compute the optical and thermal conductivities of Holstein chains at finite temperatures and filling. We validate our approach by comparison with analytical results applicable to single polarons valid in the small polaron limit. Our work provides a systematic study of contributions to the thermal conductivity at finite frequencies and elucidates differences in the spectrum compared to the optical conductivity, covering both the Luttinger-liquid and charge-density-wave regimes. Finally, we demonstrate that our approach is capable of extracting the DC conductivities as well. Beyond this first application, several future extensions seem feasible, such as, the inclusion of dispersive phonons, different types of local electron-phonon coupling, and a systematic study of drag effects in this electron-phonon coupled system.
\end{abstract}

\maketitle
\textit{Introduction.} 
Computing the transport properties of strongly correlated quantum many-body systems with rigorous approaches remains
a key challenge in condensed matter theory. This applies even to one-dimensional (1D) systems, 
as soon as multiple local degrees of
freedom are involved, even though
 powerful analytical and numerical methods are often applicable~\cite{bertini_21,Bulchandani2021}. 
 For instance, phonons are ubiquitous in a solid-state environment and provide an obvious
relaxation channel for charge carriers. In  practically all materials, the thermal current has a contribution
from phonons, possibly accompanied by electronic \cite{Mahan} or spin-excitation transport \cite{Hess2019}.

Methodological developments could be utilized in several experimentally relevant contexts. One example concerns quasi-1D materials that undergo a Peierls transition (see e.g., Refs.~\cite{perfetti_02,li_22}, or interfaces and heterostructures of materials involving small-polaron physics, such as manganites~\cite{Schramm_2008,saucke_12}). Second, for 1D quantum magnets, a fully quantum treatment of phonons and their role in transport problems would be desirable, complementing a body of analytical work \cite{Shimshoni2003,Chernyshev2005,Boulat2007,Gangadharaiah2010,Bartsch2013,Chernyshev2015,Chernyshev2016}. Third, the thermal transport of correlated materials suggested for next-generation solar cells~\cite{kressdorf_20} poses another interesting challenge.

Common theoretical approaches to capture transport in electron-phonon coupled systems include Boltzmann theory~(see, e.g., Refs.~\cite{Chernyshev2015,Gangadharaiah2010,amarel_20,amarel_21}),
dynamical mean-field theory  in higher dimensions~\cite{fratini_01,fratini_03,fratini_06},  and in some cases also quantum Monte Carlo (QMC) simulations~\cite{Louis2003,
mishcenko_03,mishcenko_08,
mischenko_15,weber_Freericks_2021}. Applying matrix-product-states~\cite{schollwock2011density} (MPS) methods has the advantage 
that both inhomogeneous and frustrated systems can be treated and dynamical information can be obtained from a single time evolution. The main challenges consist in first, an 
efficient treatment of the phonon degrees of freedom and second, for time-dependent approaches,  
in reaching sufficiently long times. We here report the successful application of an MPS algorithm using
local basis optimization (LBO) \cite{zhang98,Guo2012,brockt_dorfner_15,stolpp2020,jansen20,jansen22} of the phonon state space combined with finite-temperature techniques \cite{verstrate2004,barthel2009,barthel2012,barthel2013,
karrasch2012,karrasch2013,kennes2016,barthel_16} and state-of-the-art 
time-evolution schemes \cite{paeckel_2019} to obtain the optical and thermal conductivity of the paradigmatic Holstein chain 
at a finite electronic filling, with a  focus  on finite-frequency properties. 

We stress that there is significant information in the finite-frequency data, for both charge
and thermal transport.
 For example, optical conductivity data obtained from absorption spectra has been used to verify the presence of small polarons in manganites~\cite{Quijad_98,hartinger_06,mildner_15}.  Optical spectroscopy was also used to study the formation of a charge density wave (CDW) gap when the temperature is lowered, see, e.g., Ref.~\cite{li_22}. Furthermore, the development of methods such as the 3$\omega$ method~\cite{cahill_87} and time-domain and frequency-domain thermoreflectance~\cite{paddock_86,schmidt_09} enhanced the demand for accurate theoretical descriptions of the physical processes that contribute to the thermal conductivity.

The Holstein model captures local electron-phonon interactions~\cite{Holstein1959} with optical phonons. While seemingly simple, the complexity induced by this interaction has made it an interesting model for studying both polaron and CDW formation~\cite{Holstein1959_2,hirsch_83,
bursill_98,creffield_05,franchini_21}. 
 For example, the Holstein polaron and the finite-filling optical conductivities have been actively investigated in, e.g., Refs.~\cite{capone_97,zhang99,fratini_01,schubert_05,
fratini_06,Loos2007,wellein98,goodvin11,weber_16,
bonca_21,fratini_21,jansen22}, and the polaronic contribution to energy transport in the Holstein model was investigated via a Green's function approach in the static limit in Ref.~\cite{rezania_12}. In Ref.~\cite{weber_assaad_18}, Weber \textit{et al.}~used a QMC method to study the compressibility, specific heat, and spectral functions for a variety of parameters in the one-dimensional Holstein chain.

In this Letter, we compute the real part of the optical and thermal conductivity of the Holstein model at half-filling and at finite temperatures 
in both the Luttinger-liquid (LL) and CDW regime.
We elucidate the difference between the two regimes and study the frequency dependence starting from the well-controlled limit of small polarons.
We  demonstrate that our approach also allows for extracting the DC optical conductivity.

\textit{Model.} The spinless Holstein model~\cite{Holstein1959} describes spinless fermions propagating with a hopping amplitude $t_0$,  interacting with local harmonic oscillators with a frequency $\omega_0$. The electron-phonon  coupling strength  is given by $\gamma$. The Hamiltonian for an $L$ site system with open boundary conditions  and $\hbar=1$ reads 
\begin{equation} \label{eq:def_HolHam}
  \hat H=- t_0\sum\limits_{j=1}^{L-1} \left( \hat c_{j}^{\dag} \hat c_{j+1}^{\phantom{\dag}} +\textrm{H.c.}\right) +\sum\limits_{j=1}^L \bigl(\omega_0 
\hat n_j^b +\gamma (\hat n_j-\frac{1}{2})  \hat{X}_j\bigr)\, ,
\end{equation}
with $\hat c^{\dag}_{j}$ $ (\hat b^{\dag}_{j})$ being the electron (bosonic) creation operator, and $\hat c_{j} $ $(\hat b_{j})$  the corresponding annihilation operator. Further, $\hat{X_j}=\hat b_j^{\dag} + \hat b_{j}^{\phantom{\dag}}  $ and $\hat{n}_j=\hat c^{\dag}_{j} \hat c_{j} $, $\hat{n}_j^b=\hat b^{\dag}_{j} \hat b_{j} $. We truncate the local phonon Hilbert space by allowing for at maximum $M$ phonons per site and optimize the local state space using LBO (see the Supplementary Material \cite{supp-mat}). We choose representative  parameter sets in the CDW and in the LL regime of the model:
$\gamma/t_0=\sqrt{1.6}$, $\omega_0/t_0=0.4$, and $M=40$ (CDW regime) and  $\gamma/t_0=1$, $\omega_0/t_0=1$, and $M=20$ (LL regime). 
We choose the chemical potential such that the system is particle-hole symmetric.

Our main objective is to calculate 
the charge and energy-transport coefficients, each related to a conserved charge $\hat Q=\sum_{i=1}^L \hat q_i$ with $[\hat H, \hat Q]=0$. 
To obtain the frequency-dependency of the  conductivities, we  evaluate Kubo formulas \cite{kubo57,Luttinger_64,pottier_2010,bertini_21}. 
To that end,  we calculate the time-dependent current-current correlation functions:
          \begin{equation} \label{eq:def_corel0}
C^{Q}(t)= \expval*{\hat J^{Q}(t)\hat J^{Q}(0)}_T \, ,
\end{equation} where the subindex $T$ indicates a thermal expectation value in the canonical ensemble at temperature $T$.  Furthermore,  
we use the label $Q=E(C)$ for the energy(charge) current.  The charge current for the Holstein model is
    \begin{equation} \label{eq:def_Ccurr}
\hat J^{C}=it_0 \sum\limits_{i=1}^{L-1}(\hat c_{j}^{\dag} \hat c_{j+1}^{\phantom{\dag}} - \hat c_{j+1}^{\dag} \hat c_{j}^{\phantom{\dag}} ) \, ,
\end{equation}
and the energy current becomes~\cite{schoenle_21}
  \begin{equation} \label{eq:def_Ecurr}
\begin{split}
\hat J^{E} =
 \hat J^E_e + \hat J^E_{e-ph}\, ,
\end{split}
\end{equation}
with the two contributions 
  \begin{eqnarray} 
 \hat J^E_e &=&it_0^2 \sum\limits_{j=2}^{L-1} (\hat c^{\dagger}_{j-1} \hat c_{j+1}- \hat c^{\dagger}_{j+1}c_{j-1})\, ,
\label{eq:def_Ee} \\
 \hat J^E_{e-ph} &=&-it_0 \gamma   \sum\limits_{j=2}^{L}(\hat c^{\dagger}_{j-1} \hat c_{j}- \hat c^{\dagger}_{j}\hat c_{j-1} )(\hat b_j+\hat b^{\dagger}_j)\, .
   \label{eq:def_Eeph}
\end{eqnarray} 
The Fourier-transformed correlation function is
   \begin{equation} \label{eq:def_corelFT}
\tilde{C}^{Q}(\omega )= \int_{-\infty}^{\infty} e^{i \omega t} f(t)C^{Q}(t)  dt \, ,
\end{equation}
with Gaussian broadening $f(t)=e^{-\abs{t}^2\eta}$.  Lastly, we obtain 
the transport coefficient $ \mathcal{L}_Q(\omega) = \mathcal{L}_Q^{\prime}(\omega)+i \mathcal{L}_Q^{\prime \prime}(\omega)$, whose 
        real part is given by
                        \begin{equation} \label{eq:def_realpart}
        \mathcal{L}_Q^{\prime}(\omega)= \frac{1-e^{-\omega /T }}{2\omega T^{\alpha}}\tilde{C}^Q(\omega) \,  ,
        \end{equation} 
         where $\alpha=0$ $(1)$ for the optical (thermal) conductivity, $ \mathcal{L}_C^{\prime}(\omega)\equiv\sigma^{\prime}(\omega)$ $ [\mathcal{L}_E^{\prime}(\omega)\equiv\kappa^{\prime}(\omega)] $.
Due to our choice of chemical potential, the thermoelectric coupling between energy and charge current vanishes, and hence the 
thermal current can be replaced by the energy current. 
Our calculation is canonical in the sector that would dominate at chemical potential $\mu=0$.
      
From our time-dependent approach, the DC conductivities can only be extracted if the current autocorrelations decay sufficiently fast and within the
accessible time range $t\leq t_{\textrm{tot}}$
(see also \cite{huang_13,steinigeweg_15} for examples for spin models).
In those cases (see below), we  obtain  the zero-frequency component  $\tilde{C}^Q(\omega=0)$ from Eq.~\eqref{eq:def_corelFT}.  Aiming at finite temperatures, we can  expand $e^{-\omega /T}\approx 1 -\frac{\omega}{T}+\mathcal{O}\bigl( (\omega/T)^2\bigr)$. Inserting this into Eq.~\eqref{eq:def_realpart} gives
                        \begin{equation} \label{eq:def_gen_dc}
                        \begin{split}
       \mathcal{L}^{\prime}_{Q,\textrm{DC}}\approx \frac{1}{2T^{\alpha+1}}\tilde{C}^{Q}(0) \,  .
       \end{split}
        \end{equation}
In our procedure, a residual dependence of  $\tilde{C}^{Q}(0)$ on the artificial broadening $\eta$ in Eq.~\eqref{eq:def_corelFT}
cannot be avoided, and both $\eta$ and $t_{\textrm{tot}}$ affect the accuracy of the results (see \cite{supp-mat} for a discussion of the
convergence).
In our work, we use open boundary conditions and study non-integrable systems at finite temperatures. There, a finite-size Drude weight may contribute at finite small frequencies~\cite{rigol_08}.

 Technically, the parameter $\epsilon_{\textrm{bond}, \hat J^Q}$ determines the bond dimension of the MPS after the application of   matrix-product operators to the initial state. This step occurs  before the real-time evolution which is carried out using the single-site time-dependent variational
principle algorithm~\cite{haegeman_11,haegeman_16}.  
In the data presented here, we use $\epsilon_{\textrm{bond}, \hat J^Q}=10^{-9}$ and $t_{\textrm{tot}} t_0=19$. All calculations are done with the ITensor Software Library~\cite{itensor}.

\begin{figure}[t]
 \includegraphics[width=0.99\columnwidth]{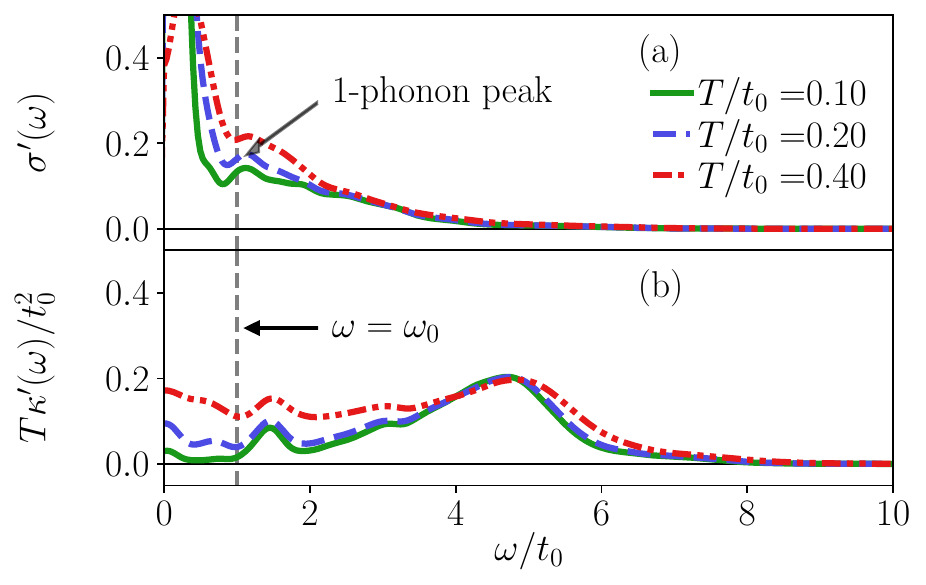}
\caption{(a) Real part of the optical conductivity for the LL parameters at different temperatures. The arrow illustrates the one-phonon emission peak. (b) Real part of the thermal conductivity for the same parameters. The vertical line shows $\omega =\omega_0$. }
\label{fig:OPCKAPMET}
\end{figure}

\textit{Results: LL regime.} 
We  work with  $L=22$ unless stated otherwise, and the Fourier transformations are done with $\eta=0.1/(4\pi)$. In Fig.~\ref{fig:OPCKAPMET}(a), we show the optical conductivity for the model at different temperatures. There, one sees a dominant peak at low frequencies, which still has a significant $L$ and $\eta $ dependence indicative of Drude-weight contributions and/or slowly decaying current correlations (not illustrated here). 
 Other features seen in the data are inherited  from the single-polaron spectra \cite{fratini_06,jansen22,bonca_21}. 
Most prominently, the single-phonon emission peak starts at $\omega \approx \omega_0$ (gray dashed vertical line). Furthermore, at large $\omega/t_0 $, we observe a temperature-independent decay. 

Figure~\ref{fig:OPCKAPMET}(b) shows the thermal conductivity for the same parameters as Fig.~\ref{fig:OPCKAPMET}(a). Here, a different picture emerges.  At low temperatures, the spectra exhibit a maximum at high frequencies, $\omega/t_0\approx 5$, which then rapidly decreases. Since the optical conductivity has practically  
decayed to zero at these values of $\omega$, the behavior of $\kappa^{\prime}(\omega)$ results from  non-vanishing energy-current matrix elements at these energies not 
present for the charge current.
 As these frequencies are larger than both the free-fermion bandwidth $4t_0$ and the phonon frequency, the spectral contribution can be attributed to multi-phonon processes. As temperature increases, we observe enhanced spectral weight at lower frequencies, corresponding to a thermal activation of transitions. The thermal conductivity also features a peak beginning at $\omega \approx \omega_0$, similar to the one-phonon emission peak in the optical conductivity.

\begin{figure}[t]
 \includegraphics[width=0.99\columnwidth]{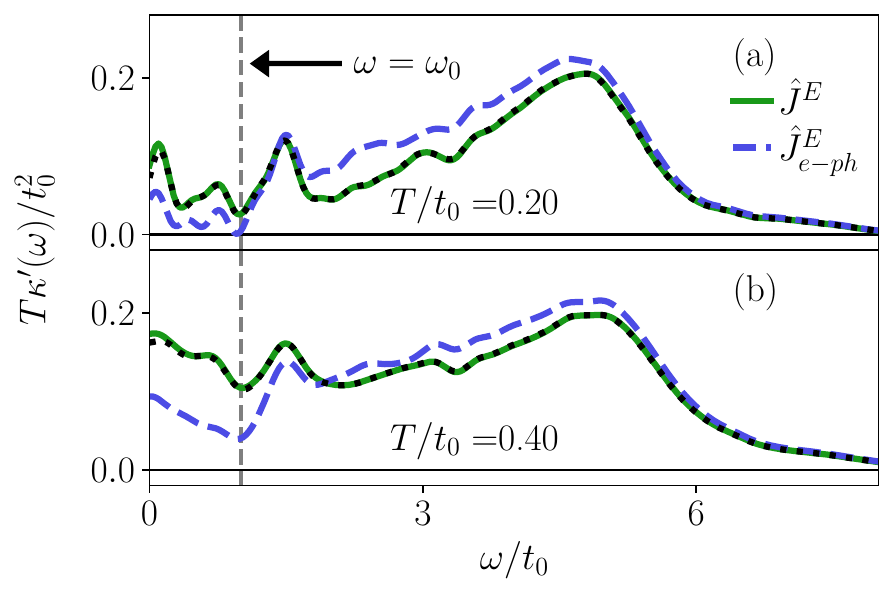}
\caption{(a) Real part of the thermal conductivity for the LL regime at $T/t_0=0.2$. (b) Same as (a) but at $T/t_0=0.4$. The solid lines are calculated using $\hat J^E$ from Eq.~\eqref{eq:def_Ecurr} and the dashed lines using $\hat J^E_{e-ph}$ from Eq.~\eqref{eq:def_Eeph}. The black dotted line shows calculations using $\hat J^E$ but with $L=20$.  The vertical line shows $\omega =\omega_0$.  }
\label{fig:KAPREDMET}
\end{figure}

To better understand the origin of the different parts of the thermal conductivity spectra, we  compute the spectra  associated with  $ \hat J^E_{e-ph}$ from Eq.~\eqref{eq:def_Eeph}.
A comparison of the results using the full current $\hat J^{E}$ (solid lines) and $\hat J^E_{e-ph}$ (dashed lines) is illustrated  for two different temperatures in Figs.~\ref{fig:KAPREDMET}(a) and ~\ref{fig:KAPREDMET}(b). In both cases, the curves almost overlap at high frequencies, 
implying that the high-frequency structures result from the term in $\hat J^{E}$ proportional to $\gamma$, that is, $\hat J^E_{e-ph}$. 
We attribute this to the high-energy processes induced by the coupling to the phonons, whereas the optical conductivity is limited by the number of phonons present (at low $T/\omega_0$, the spectra are dominated by phonon-emission processes). We also see that the resonance starting at $\omega /t_0=1$ can be completely attributed to the $\hat J^E_{e-ph}$ term. The black dashed line in Fig.~\ref{fig:KAPREDMET} is calculated with $L=20$, and apart from very low frequencies, the spectra are almost indistinguishable.  

At small $\omega/t_0$, the spectra associated with $\hat J^E$ and  $\hat J^E_{e-ph}$ start to deviate from each other, which signals that $\hat J^E_{e}$  starts to play a more important role. This also becomes more prominent at higher temperatures, and we attribute a significant portion  of the DC thermal conductivity to  $\hat J^E_{e}$ \cite{supp-mat}.

\begin{figure}[t]
 \includegraphics[width=0.99\columnwidth]{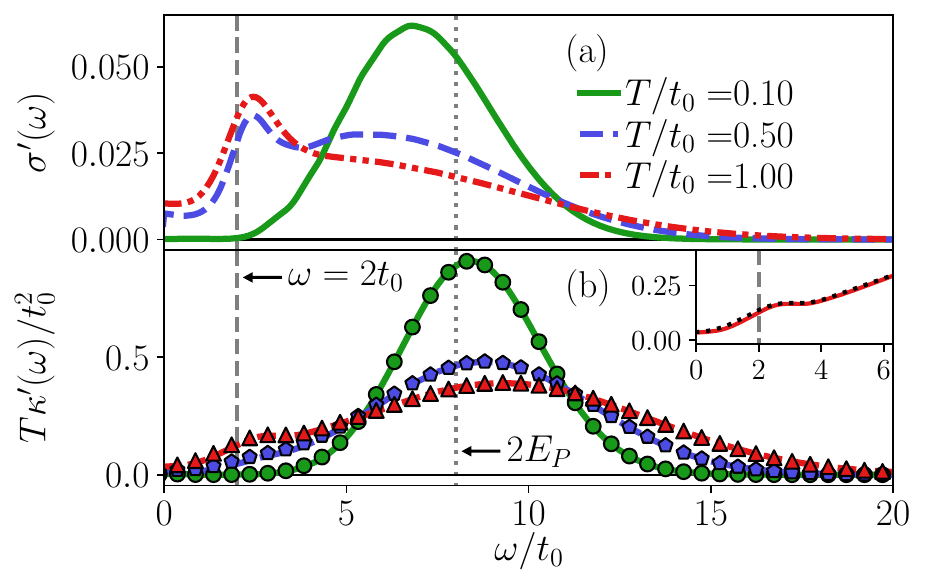}
\caption{(a) Real part of the optical conductivity for the CDW regime at different temperatures. (b) Real part of the thermal conductivity for the same parameters. The symbols in (b) show the thermal conductivity calculated using $\hat J^E_{e-ph} $ from Eq.~\eqref{eq:def_Eeph}. We only plot every 15th point of the $\hat J^E_{e-ph} $ data for clarity. The dashed vertical line illustrates $\omega=2t_0$ and the dotted vertical line $\omega=2E_P$. The inset in (b) shows the data for $T/t_0=1.0$ as a red solid line for $\hat J^E$ and a black dotted line for $\hat J^E_{e-ph}$. }
\label{fig:OPCKAPCDW}
\end{figure}

  \textit{Results: CDW regime.}
We use $L=10$ unless stated otherwise, and the Fourier transformation is done with $\eta=0.4/(4\pi)$. 
The corresponding optical conductivities are depicted in Fig.~\ref{fig:OPCKAPCDW}(a). For the polaron, it is well established (see, Refs.~\cite{Reik_67,emin_93,schubert_05,fratini_06,jansen22}), that the optical conductivity spectra are close to an asymmetric Gaussian  centered around $2E_P=2\gamma^2/\omega_0$. For the Holstein dimer ($L=2$), this can be interpreted as a transition from the lower to the excited Born-Oppenheimer (BO) surface~\cite{Born_27,Born_54} (BO  surfaces are shown in \cite{supp-mat}) at a fixed phonon configuration, however, the qualitative picture remains valid for larger systems as well, see, e.g., Refs.~\cite{jansen22,schubert_05,fratini_06}.  Furthermore, a thermally activated transition for the polaron occurs at $\omega\approx 2 t_0$ \cite{schubert_05}, which can be explained using the BO surfaces as a transition at $\expval*{\hat X_1-\hat X_2}\approx 0$.

 In Fig.~\ref{fig:OPCKAPCDW}(a), we see that for the selected  parameters, the polaron properties carry over to half filling. The center of the spectrum is close to $2E_P$ (dotted gray lines) and the thermally activated resonance is clearly visible (dashed line). Note that this peak is also seen for classical phonons, studied in Ref.~\cite{weber_16}, and has been suggested to be related to finite-temperature disorder physics of noninteracting electrons. Furthermore, it was also discussed in the context of the displaced Drude peak in Ref.~\cite{fratini_21}.

The thermal conductivity calculated with $\hat J^E_{e-ph}$ and $\hat J^E$ is shown in Fig.~\ref{fig:OPCKAPCDW}(b).  The spectra look very similar to those of the optical conductivity, suggesting  that they are dominated by small-polaron physics. For the Holstein dimer, $\hat J^E= \hat J^E_{e-ph}$, and in the semi-classical approach underlying the BO picture where the phonon states are eigenstates of $\hat X_i$, one would just expect a rescaled optical conductivity. Our data indicate that the picture is qualitatively correct, but with some notable differences. 
For example, the center of the spectra is shifted to $\omega >2E_P$, and the thermally activated resonance appears as a shoulder peak at $T/t_0=1.0$ [inset in Fig.~\ref{fig:OPCKAPCDW}(b)]. The data presented in this figure establish  that the spectrum mainly results from  $\hat J^E_{e-ph}$ (symbols) for all frequencies visible on the used scale. 
Thus, the main features can be traced to  Holstein-dimer physics.

\begin{figure}[t]
\includegraphics[width=0.99\columnwidth]{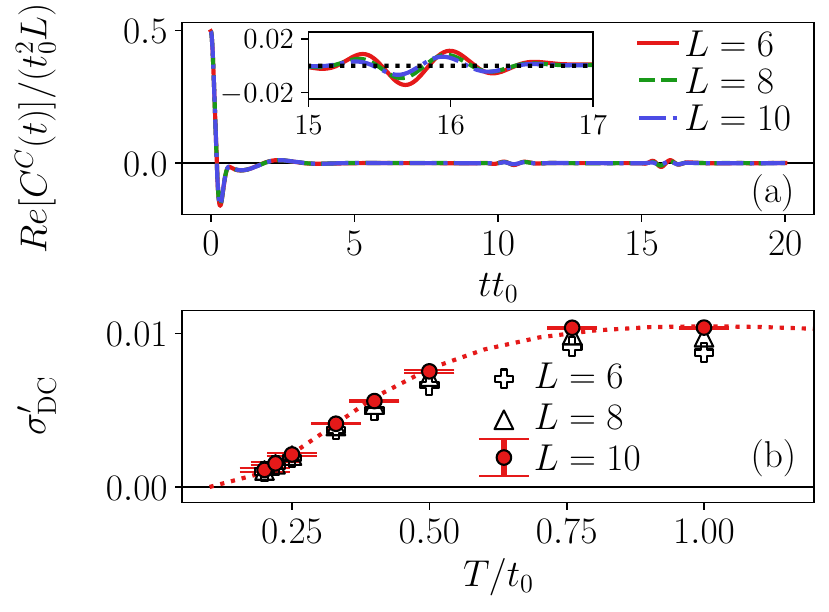}
\caption{(a) Time-dependent charge current-current correlation functions for the CDW parameters at $T/t_0=1.0$ and different system sizes $L$. The inset shows the oscillations at times $15\leq t t_0 \leq 17$ on a smaller scale. The black dotted line in the inset shows the data weighted with $e^{-\eta (tt_0)^2}$ with $\eta=0.4/(4\pi) $. (b) DC optical conductivity calculated for the same parameters via Eq.~\eqref{eq:def_gen_dc}. The error bars are calculated by varying $\eta$ (see the main text for details). The dotted line is a fit of $\rho_P^{-1}$ from Eq.~\eqref{eq:def_rhopol} to the data.}
\label{fig:CDWDC}
\end{figure}

We next demonstrate that our data can be used to extract DC quantities. In the main text, we focus on $\sigma^{\prime}_{\textrm{DC}}$, while results for $\kappa^{\prime}_{\textrm{DC}}$ are presented in \cite{supp-mat}. 
We first demonstrate that for the selected parameters, the time-dependent correlation function quickly decays to zero [see Fig.~\ref{fig:CDWDC}(a)].
  Some oscillations can be seen at later times, but they decrease with system size, and $\eta $ is chosen such  that the conductivity should be independent thereof [the black dashed line in the inset in Fig.~\ref{fig:CDWDC}(a) shows the correlation function weighted with $e^{-\eta (tt_0)^2 }$].
$\sigma^{\prime}_{\textrm{DC}}$ is then obtained as the zeroth component of the Fourier transformed charge current-current correlation function, see Eq.~\eqref{eq:def_gen_dc}.  

 In systems with small polarons, $\sigma^{\prime}_{\textrm{DC}}$ increases due to thermally activated hopping. As demonstrated in Fig.~\ref{fig:CDWDC}(b), this picture remains valid at half filling. For small polarons, an analytical expression for the resistivity can be derived \cite{Bogomolov_69,emin_69,austin_01}

\begin{equation} \label{eq:def_rhopol}
\rho_P=\rho_0Te^{(0.5 E_P-t_0)/T} \, .
\end{equation}
In Fig.~\ref{fig:CDWDC}(b), we show $\sigma^{\prime}_{\textrm{DC}}$ as a function of temperature. By fitting the inverse of Eq.~\eqref{eq:def_rhopol} with $\rho_0$ as the fitting parameter to our data, we obtain a very good agreement, and further confirm the importance of small-polaron transport in the system, even at half filling. At large temperatures,  
the residual  finite-size effects are consistent with the remaining $L$-dependence in the autocorrelators [see the inset of Fig.~\ref{fig:CDWDC}(a)]. 
Note that the error bars in Fig.~\ref{fig:CDWDC}(b) result from  comparing data with  $\eta=0.2/(4 \pi)$ and $\eta=0.6/(4 \pi)$ and taking the maximum difference to the $\eta=0.4/(4 \pi)$ data at each point.
By comparison to state-of-the-art numerical results for one-dimensional spin systems \cite{huang_13,steinigeweg_15}, we conclude that our method
reaches comparably low temperatures yet for much larger local Hilbert spaces.

\textit{Conclusions.} In this letter, we studied the thermal and optical conductivity for the Holstein model at finite temperature and half-filling in the  Luttinger-liquid  and in the  CDW regime. Using state-of-the-art MPS time-evolution techniques and using  LBO to efficiently deal with the phonons, 
we computed the energy and charge current-current correlation functions. From those functions, we extracted the finite-frequency transport coefficients. Despite the enhanced complexity of considering finite filling and the complicated energy-current operator, we succeeded in obtaining results at low-temperatures and investigated 
 the role of different terms of the energy-current operator  in the two parameter regimes. In particular, our data reveal the importance of small polarons as energy carriers for parameters with a CDW ground state at these temperatures. We further provided an example where the DC conductivity can be reliably obtained in the CDW regime and we identified features in the spectra due to multiphonon processes.

While the spectra of the optical and the thermal conductivity share some similarities, in particular, in the small-polaron regime, there are also
noteworthy differences. 
In the LL regime, the thermal conductivity acquires a high-frequency feature, not seen for $\sigma^\prime(\omega)$, that we attribute to multi-phonon
processes.
In the CDW regime, the thermal conductivity spectrum has a much larger amplitude, the center of the spectra is shifted to higher frequencies, and the thermally activated resonance can only be distinguished at higher $T/\omega_0$. Still, the resemblance to the optical conductivity can be understood by looking at the energy-current operator in the dimer limit. 
Moreover, we unvelied the contribution of small-polaron physics to the   optical conductivity (both at finite and zero frequency) and, in particular, to the thermal conductivity in the CDW regime.

One can directly identify  several interesting extensions of our work. First, one would want to systematically study the DC conductivity starting from the limit $\gamma=0$, accounting also for
a dispersion of the phonons and non-nonlinearity in the phonon sector.
Second, the method can be applied to other models as well \cite{szabo_21}. For instance, transport of a Heisenberg chain coupled to phonons has not been explored with MPS methods 
(see \cite{Louis2003, Louis2006} for related numerical studies).  This might provide new insights into experiments on Heisenberg-chain-like systems~\cite{Sologubenko_07,Hess2019}.  A different direction would be to extend the analysis  to non-equilibrium quantities, see, e.g., Refs.~\cite{sentef13,kemper_17,paeckel_fauseweh_19,rincon_21,ejima_22}, which might require additional sophisticated matrix-product-based schemes~\cite{jeckelmann98,schroeder_16,bischoff_17,
yang_20,koehler20,Xu_21,mardazad_21,moroder_22}.

The data shown in this manuscript are available as ancillary files. The files and a link to the code used can be found on the arXiv.

\textit{Acknowledgments.}
 We acknowledge useful discussions with J. Bon\u{c}a, E. Jeckelmann, and  J. L\"{o}tfering, and we thank M. Weber for comments on a previous version of the manuscript.
This work is funded by the Deutsche Forschungsgemeinschaft (DFG, German Research Foundation) – 217133147, 436382789, 493420525  via SFB 1073 (project B09) and  large-equipment grants (GOEGrid cluster). 
D.J. also acknowledges the support from the Government of Spain (European
Union NextGenerationEU PRTR-C17.I1, Severo Ochoa CEX2019-000910-S and TRANQI), Fundació Cellex, Fundació Mir-Puig, and Generalitat de Catalunya (CERCA program).
\beginsupplement
\section*{Supplementary material}
\label{sec:sup}
\begin{figure}[t]
\includegraphics[width=0.99\columnwidth]{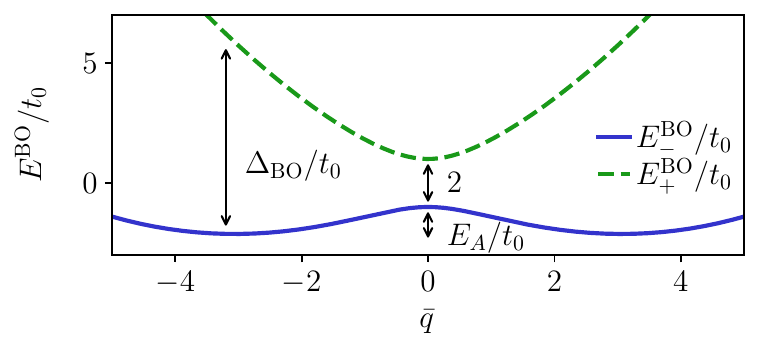}
\caption{Born-Oppenheimer surface for the CDW parameters. $E_A$ is the difference between the minimum and maximum energy of the $E^{\textrm{BO}}_-$. }
\label{fig:BOsurf}
\end{figure}

\textit{Born-Oppenheimer surfaces.} For the charge-density-wave (CDW) parameters, some of our results can be understood based on the Born-Oppenheimer Hamiltonian and the Born-Huang formalism~\cite{Born_27,Born_54}. For the Holstein model in the dimer limit, see Ref.~\cite{jansen22} for the convention we use, the Born-Oppenheimer Hamiltonian becomes
\begin{equation} \label{eq:def_BO}
\hat H^{\rm{BO}}=-t_0 \bigl( \hat c_{1}^{\dag} \hat c_{2}^{\phantom{\dag}} +\textrm{H.c.}\bigr)+\gamma \bigl[\bar{q}(\hat n_1 - \hat n_2)\bigr]+\frac{1}{2}\bar{q}^{2}\omega_0\, ,
\end{equation}
which has the eigenenergies 
\begin{equation} \label{eq:def_BO_surfaces}
E^{\rm{BO}}_{\pm}=\frac{1}{2}\Bigl ( \bar{q}^{2}\omega_0 \pm 2\sqrt{(\gamma\bar{q})^{2}+t_0^2} \Bigr) \, .
\end{equation}
 $E^{\rm{BO}}_{\pm}$ as a function of $\bar q$ are plotted in Fig.~\ref{fig:BOsurf}. For our parameters, $\Delta^{\rm{BO}}\approx 2E_P$ [up to order $\mathcal{O}(10^{-4})$]. The activation energy $E_{A}/t_0=1.125>0.5E_P/t_0-1=1$ gives the exponent in the DC resistivity and we find that using $E_{A}/t_0=0.5E_P/t_0-1$ gives a better fit of our DC conductivity data to $\rho_P$ from Eq.~(10) of the main text (relative error for the fitting parameter is $0.01$ instead of $0.03$). At $\bar{q}=0$ $\Bigl(\expval*{\hat X_1-\hat X_2}=0\Bigr)$, we also see the thermally activated transition with an energy difference $\omega / t_0=2$.

\textit{Convergence of the algorithm.} In this work, we calculate the time-dependent correlation functions at finite temperatures using matrix-product states (MPSs) with purification~\cite{verstrate2004,schollwock2011density}. For the imaginary time-evolution, we use the two-site time-dependent variational principle  (TDVP) algorithm~\cite{haegeman_11,haegeman_16} with local basis optimization (LBO)~\cite{zhang98}. 
   \begin{figure}[t]
\includegraphics[width=0.99\columnwidth]{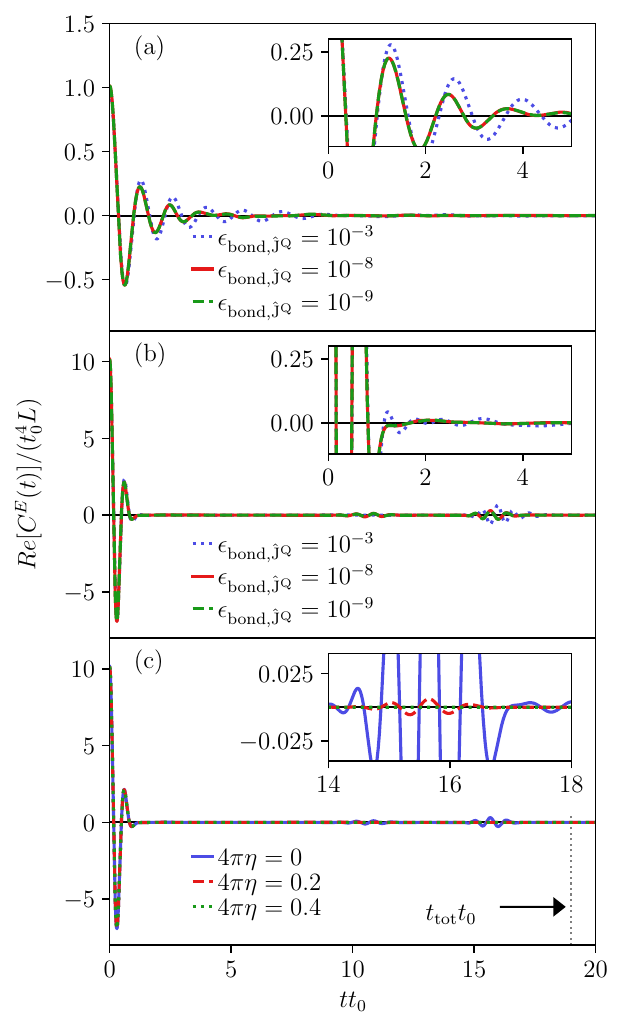}
\caption{(a) Energy current-current correlation function for the LL parameters, $T/t_0=0.4$, and different $\epsilon_{\textrm{bond}, \hat J^Q}$ [see Eq.~\eqref{eq:def_mpo_trunc}]. (b) Same as (a) but for CDW parameters, $T/t_0=0.5$, and $L=10$. (c) Same as (b) but using $\epsilon_{\textrm{bond}, \hat J^Q}=10^{-9}$ and weighting the data with $e^{-\eta (tt_0)^2}$ using different $\eta$. The insets show the corresponding data on a smaller scale. The dotted vertical line in (c) illustrates $t_{\textrm{tot}}t_0=19.0$. }
\label{fig:diffeps}
\end{figure}
When we carry out the singular-value decompositions, we truncate based on the singular values $s_{\eta}$ such that 
\begin{equation} \label{eq:def_svd_trunc}
\sum\limits_{\textrm{discarded} \, \eta}s^2_{\eta}/(\sum\limits_{\textrm{all} \, \eta}s^2_{\eta}) <\epsilon_{\textrm{bond}}\, .
\end{equation}
When we compute the optimal basis in LBO, we truncate the eigenvalues of the reduced density matrix, $w_{\eta}$, such that
\begin{equation} \label{eq:def_lbo_trunc}
\sum\limits_{\textrm{discarded}\,  \eta}w_{\eta}/(\sum\limits_{\textrm{all} \, \eta}w_{\eta}) <\epsilon_{\textrm{LBO}}\, .
\end{equation}
For all the data shown in this work, we use $\epsilon_{\textrm{LBO}}=\epsilon_{\textrm{bond}}=10^{-9}$ and an imaginary time step $d \tau t_0=0.1$. To further verify that the data are converged with respect to varying those parameters, we calculate $\expval*{[\hat A^{Q}, \hat J^{Q}]}_T$ [related to the conductivities via the $f$-sum rule in Eq.~\eqref{eq:def_fsum}] for $Q=C$ and $Q=E$, $\expval*{\hat H}$, and $\expval*{\hat H^2} $ (the latter two are needed to compute the specific heat of the system). For the Luttinger-liquid (LL) parameters, we found a maximum relative difference for the observable $\expval*{[\hat A^{E}, \hat J^{E}]}_T$ of $0.017$ when using $d \tau t_0=0.01$ instead of $d \tau t_0=0.1$ for the temperatures used in this work (the maximum difference is for $T/t_0=0.1$ and decreasing when $T/t_0$ is increased). For the CDW parameters, the maximum relative difference is found for the same observable, and is $0.0062$ for the same change in $d\tau t_0$ at $T/t_0=1/3$.

During the real-time evolution, we use single-site TDVP~\cite{haegeman_11,haegeman_16} and evolve the component of the state in the physical Hilbert space forwards in time and the component in the ancillary Hilbert space backward in time~\cite{karrasch2012,karrasch2013,kennes2016} 
(note that improvements to single-site TDVP have been suggested, e.g., Refs.~\cite{yang_20,Xu_21}). 
Since the algorithm does not increase the bond dimension of the state, we make sure that the dimension is sufficiently large by adapting the truncation parameter of the matrix-product state after applying the current operators $\hat J^Q$ as matrix-product operators (MPOs). The truncation is based on the singular values such that 
\begin{equation} \label{eq:def_mpo_trunc}
\sum\limits_{\textrm{discarded} \, \eta}s^2_{\eta}/(\sum\limits_{\textrm{all} \, \eta}s^2_{\eta}) <\epsilon_{\textrm{bond}, \hat J^Q}\, .
\end{equation} 
In Fig.~\ref{fig:diffeps}, we illustrate how the data converges by varying $\epsilon_{\textrm{bond}, \hat J^Q}$ (Fig.~\ref{fig:diffeps}(a) for the LL parameters and Fig.~\ref{fig:diffeps}(b) for the CDW parameters). When $\epsilon_{\textrm{bond}, \hat J^Q}$ is chosen too small, deviations in the correlation functions can be seen. Figure~\ref{fig:diffeps}(c) shows a correlation function in the CDW regime using different $\eta$. The initial decay is approximately independent thereof, but the later oscillations are significantly damped. We note that in general, convergence of both the correlation functions and the transport coefficients themselves must be monitored carefully. 

\begin{figure}[t]            
    \centering              
    \def\svgwidth{250pt}    

\begingroup%
  \makeatletter%
  \providecommand\color[2][]{%
    \errmessage{(Inkscape) Color is used for the text in Inkscape, but the package 'color.sty' is not loaded}%
    \renewcommand\color[2][]{}%
  }%
  \providecommand\transparent[1]{%
    \errmessage{(Inkscape) Transparency is used (non-zero) for the text in Inkscape, but the package 'transparent.sty' is not loaded}%
    \renewcommand\transparent[1]{}%
  }%
  \providecommand\rotatebox[2]{#2}%
  \newcommand*\fsize{\dimexpr\f@size pt\relax}%
  \newcommand*\lineheight[1]{\fontsize{\fsize}{#1\fsize}\selectfont}%
  \ifx\svgwidth\undefined%
    \setlength{\unitlength}{368.09092673bp}%
    \ifx\svgscale\undefined%
      \relax%
    \else%
      \setlength{\unitlength}{\unitlength * \real{\svgscale}}%
    \fi%
  \else%
    \setlength{\unitlength}{\svgwidth}%
  \fi%
  \global\let\svgwidth\undefined%
  \global\let\svgscale\undefined%
  \makeatother%
  \begin{picture}(1,0.19914622)%
    \lineheight{1}%
    \setlength\tabcolsep{0pt}%
    \put(0,0){\includegraphics[width=\unitlength,page=1]{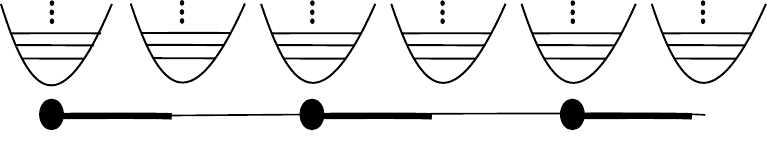}}%
    \put(0.13947388,0.00104769){\color[rgb]{0,0,0}\makebox(0,0)[lt]{\lineheight{1.25}\smash{\begin{tabular}[t]{l}1\end{tabular}}}}%
    \put(0.31037766,0.00054467){\color[rgb]{0,0,0}\makebox(0,0)[lt]{\lineheight{1.25}\smash{\begin{tabular}[t]{l}2\end{tabular}}}}%
    \put(0.48018457,0.00054467){\color[rgb]{0,0,0}\makebox(0,0)[lt]{\lineheight{1.25}\smash{\begin{tabular}[t]{l}3\end{tabular}}}}%
    \put(0.65072255,0.00105169){\color[rgb]{0,0,0}\makebox(0,0)[lt]{\lineheight{1.25}\smash{\begin{tabular}[t]{l}4\\\end{tabular}}}}%
    \put(0.81968131,0.00105169){\color[rgb]{0,0,0}\makebox(0,0)[lt]{\lineheight{1.25}\smash{\begin{tabular}[t]{l}5\\\end{tabular}}}}%
    \put(0,0){\includegraphics[width=\unitlength,page=2]{drawing.pdf}}%
  \end{picture}%
\endgroup%
  
    \caption{Matrix-product state consisting of physical (black) and ancillary (red) sites connected to local harmonic oscillators for $L=3$. In our implementation, each site corresponds to a separate tensor. In the purification ansatz, the physical and one ancillary site are maximally entangled at infinite temperature, as illustrated by the thick lines. The other bonds have dimension $d_{\textrm{bond}}=1$. Note that in practice, we work in the canonical ensemble, which does increase the bond dimensions in the system, see Ref.~\cite{barthel_16}. The numbers indicate the labeling of the bonds shown in Fig.~\ref{fig:eps_bonds}.}  
    \label{fig:pic}        
\end{figure}

Since the current operators only act on the physical Hilbert space, we demonstrate how all bonds are impacted by acting on the state with the MPO. We sketch our implementation of the MPS purification ansatz in Fig.~\ref{fig:pic}. We use one separate tensor with both fermionic and bosonic degrees of freedom for each site (physical or ancillary). We then maximally entangle the phononic modes before we apply fermionic creation operators following Ref.~\cite{barthel_16} to do the calculations in the canonical ensemble.

              \begin{figure}[t]
\includegraphics[width=0.99\columnwidth]{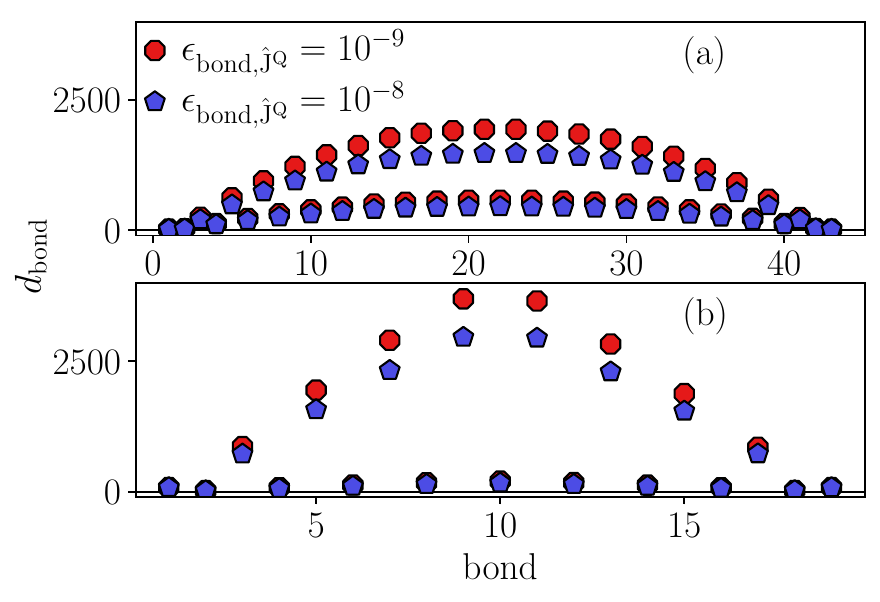}
\caption{(a) Bond dimensions of the matrix-product state after applying the energy-current operator for the LL parameters with $L=22$ and $T/t_0=0.4$, and different $\epsilon_{\textrm{bond}, \hat J^Q}$. (b) Same as (a) but for the CDW parameters with $L=10$ and $T/t_0=1.0$. Note that a sketch of the MPS ansatz can be seen in Fig.~\ref{fig:pic}.}
\label{fig:eps_bonds}
\end{figure}
             \begin{figure}[t]
\includegraphics[width=0.99\columnwidth]{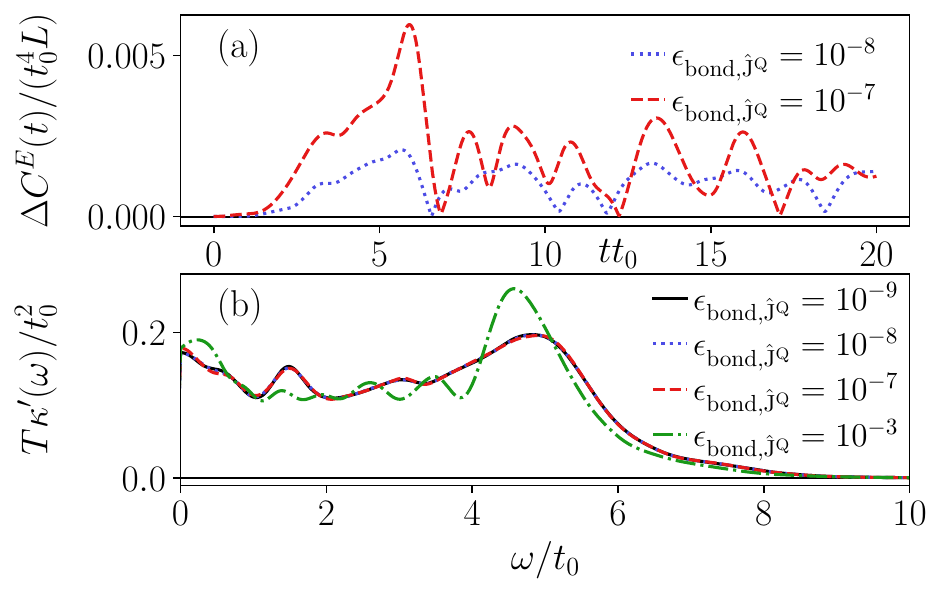}
\caption{(a) Absolute value of the difference between the energy-energy correlation function calculated with  $\epsilon_{\textrm{bond}, \hat J^Q}=10^{-8}$ and $\epsilon_{\textrm{bond}, \hat J^Q}=10^{-9}$ (blue dotted line) and with  $\epsilon_{\textrm{bond}, \hat J^Q}=10^{-7}$ and $\epsilon_{\textrm{bond}, \hat J^Q}=10^{-9}$ (red dashed) line (see Eq.~\eqref{eq_diff_C}). (b) Real part of the thermal conductivity for different $\epsilon_{\textrm{bond}, \hat J^Q}$ and $\eta=0.1/(4\pi)$. Note that the influence of $\epsilon_{\textrm{bond}, \hat J^Q}$ on the DC thermal conductivity can be seen in Fig.~\ref{fig:KAP_DC_1}. All calculations are done in the LL regime with $L=22$. }
\label{fig:diffeps_2}
\end{figure}
In Fig.~\ref{fig:eps_bonds}, we show the bond dimensions after applying the energy current operator. There, one can see that the bond dimension between the originally maximally entangled physical and ancillary sites increases when $\epsilon_{\textrm{bond}, \hat J^Q}$ is varied, especially towards the bulk of the system. The other bond dimensions (away from the edges) also increase, but the difference is not visible on the scales of the figures.

\begin{figure}[t]
\includegraphics[width=0.99\columnwidth]{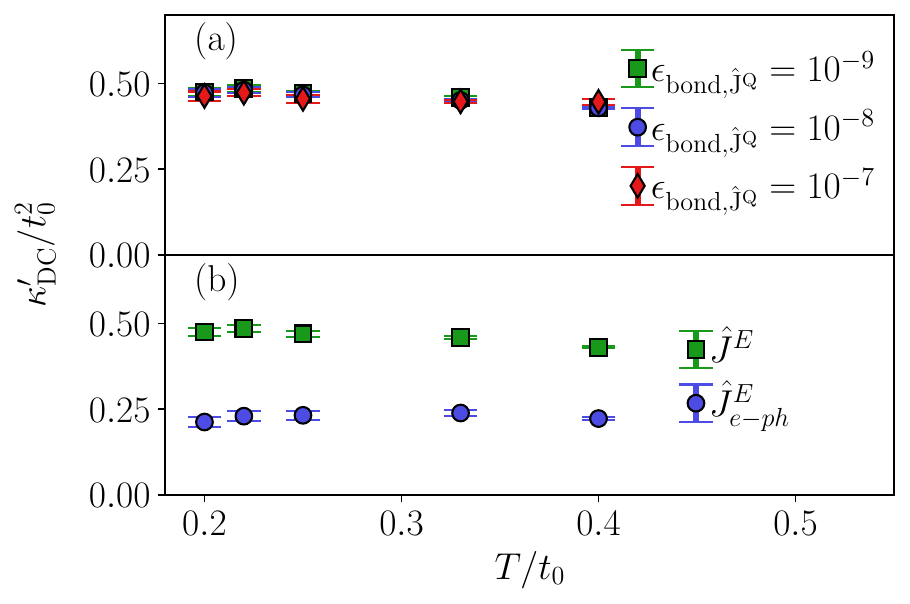}
\caption{DC thermal conductivity for the LL parameters with $L=22$ and $\eta=0.1/(4\pi)$.
(a) Using different truncation parameters $\epsilon_{\textrm{bond}, \hat J^Q}$.
(b) Comparison between using $\hat J^E$ to $\hat J^E_{e-ph}$ with $\epsilon_{\textrm{bond}, \hat J^Q}=10^{-9}$.
}
\label{fig:KAP_DC_1}
       \end{figure}

To illustrate the influence of $\epsilon_{\textrm{bond}, \hat J^Q}$ on our data, we plot 
\begin{equation}
\label{eq_diff_C}
\Delta C^E (t)=\abs{C^E_{\epsilon_{\textrm{bond}, \hat J^Q}=10^{-9}} (t)-C^E_{\epsilon_{\textrm{bond}, \hat J^Q}} (t)} \, ,
\end{equation}
where the second $\epsilon_{\textrm{bond},\hat J^Q}$ is either $10^{-7}$ or $10^{-8}$, in Fig.~\ref{fig:diffeps_2}(a). As can be seen, some differences are unavoidable, however, in Fig.~\ref{fig:diffeps_2}(b), we show how these differences affect the thermal conductivity. On the scale of interest in this work, the three curves for $\epsilon_{\textrm{bond}}=10^{-7}$, $10^{-8}$, and $10^{-9}$ are indistinguishable. We also show the data for $\epsilon_{\textrm{bond}}=10^{-3}$ to demonstrate that a too large  $\epsilon_{\textrm{bond}}$ significantly affects the data. Note that the influence on the DC conductivity can be seen in Fig.~\ref{fig:KAP_DC_1} and will be discussed later.

\textit{Fourier transformation and $f$-sum rule.}
        In this work, the real-time evolution is done up to time $t_{\textrm{tot}}t_0$ and we add $4 \cdot t_{\textrm{tot}}t_0 $ zeros to the signal (zero padding). Before the Fourier transformation, the correlation functions are also weighted with the Gaussian damping $e^{-\eta (tt_0)^2}$. To get an estimation of the accuracy of our results and to determine $\eta$, we relate the spectra to thermal expectation values.

        \begin{figure}[t]
\includegraphics[width=0.99\columnwidth]{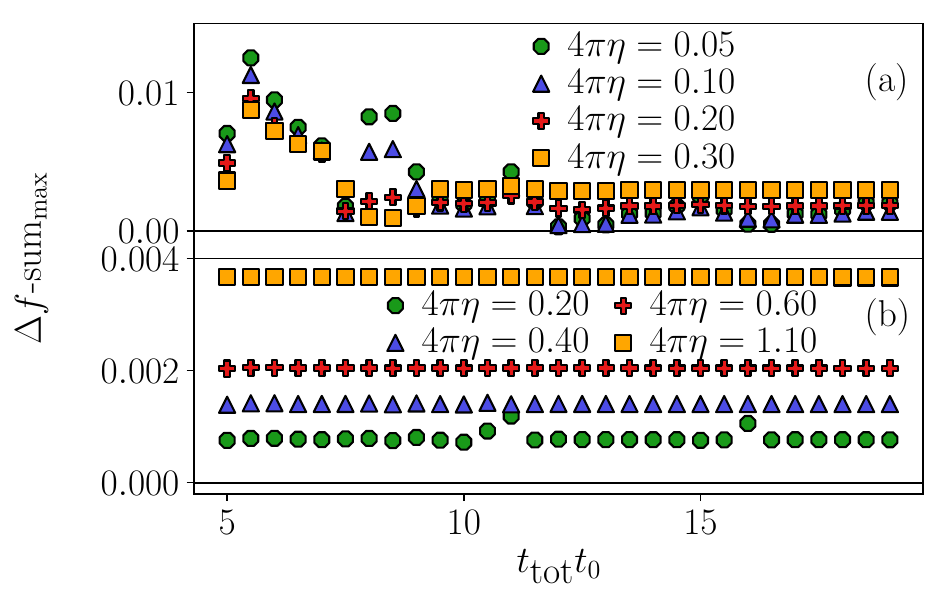}
\caption{(a) $\Delta f_{\max}$-sum for the thermal conductivity, see Eq.~\eqref{eq:def_deltaf_max}, as a function of $t_{\textrm{tot}}t_0$ for the metallic parameters with $L=22$ and using different broadening parameters $\eta$. (b) Same as (a) but for the CDW parameters and with $L=10$. }
\label{fig:fsum_rule_comp}
\end{figure}   
       \begin{figure}[t]
\includegraphics[width=0.99\columnwidth]{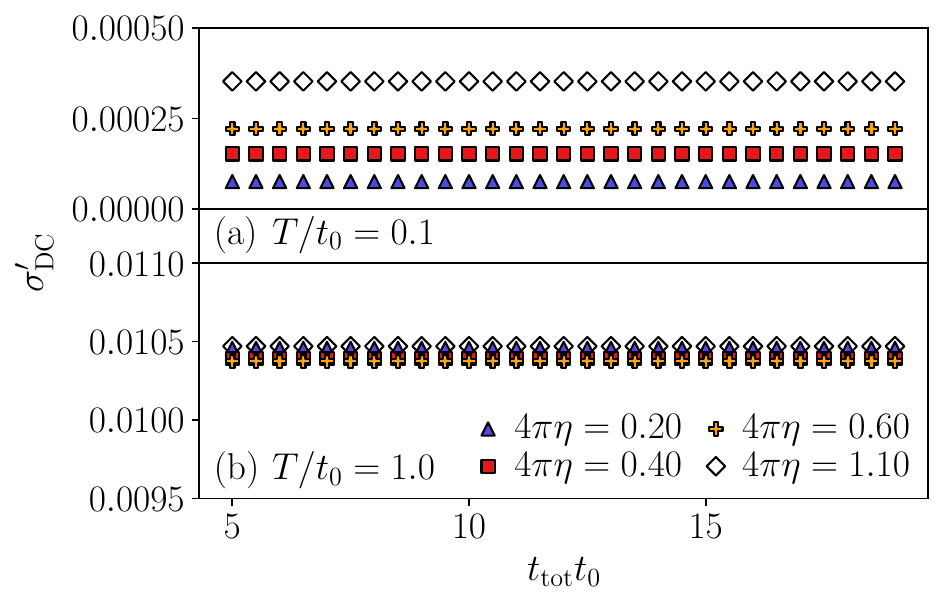}
\caption{(a) DC optical conductivity in the CDW regime with $L=10$ at $T/t_0=0.1$ as a function of $t_{\textrm{tot}}t_0$ using different $\eta$. (b) Same as (a) but at $T/t_0=1.0$. (a) Illustrates a case where the relative uncertainty is larger than 15\%, and the DC conductivity is excluded at this temperature.}
\label{fig:kappa_dc_eta}
\end{figure}

        The conductivities can be related to the commutator of the current and a polarization $\hat A^Q=\sum_{j=1}^{L}(j-1)\hat q_j$ via the $f$-sum rule
                        \begin{equation} \label{eq:def_fsum}
\begin{split}
\int_{0}^{\infty}d\omega \mathcal{L}_Q^{\prime}(\omega)=\frac{\pi}{2i}\frac{\expval*{[\hat{A}^Q, \hat{J}^{Q}]}_T}{L} \, ,
\end{split}
\end{equation} where $[\hdots] $ is the commutator. Note that for the optical conductivity, this becomes
                     \begin{equation} \label{eq:def_fsum_opc}
\begin{split}
\int_{0}^{\infty}d\omega \sigma^{\prime}(\omega)=\frac{-\pi}{2}\frac{E_{\textrm{kin}}}{L} \, ,
\end{split}
\end{equation} where $E_{\textrm{kin}}=\expval*{- t_0 \sum_{j=1}^{L-1} ( \hat c_{j}^{\dag} \hat c_{j+1}^{\phantom{\dag}} +\textrm{H.c})}_T$. Evaluating both the left and right-hand sides of Eq.~\eqref{eq:def_fsum} is then used as a consistency check for our results.
To quantify the accuracy of the data, we define the relative difference to be 
\begin{equation} \label{eq:def_deltaf}
\Delta f\textrm{-sum}= \frac{\abs*{\int_{0}^{\infty}d\omega \mathcal{L}_Q^{\prime}(\omega)-\frac{\pi}{2i}\frac{\expval*{[\hat{A}^Q, \hat{J}^{Q}]}_T}{L}}}{\abs*{\frac{\pi}{2i}\frac{\expval*{[\hat{A}^Q, \hat{J}^{Q}]}_T}{L}}}  \, ,
\end{equation} where the $\mathcal{L}_Q^{\prime}(\omega)$ are obtained at the corresponding temperature. In our calculations, we look at $T/t_0 \in \{0.1, 0.2, 0.4\}$ in the LL regime for the optical conductivity, $T/t_0 \in \{0.1, 0.2,2/9,0.25,1/3, 0.4\}$ for the thermal conductivity, and $T/t_0 \in \{0.1, 0.2, 0.4, 0.5, 1.0\}$ in the CDW regime.

 For the thermal conductivity in the LL regime with $\eta=0.1/(4\pi)$ (shown in the main text), we find that $\Delta f\textrm{-sum}$ is at maximum of order $\mathcal{O}(10^{-3})$ for $T/t_0=0.1$ and $T/t_0=1/3$ and $\mathcal{O}(10^{-4})$ for all other temperatures. For the optical conductivity, the relative accuracy is maximum order $\mathcal{O}(10^{-3})$ for all temperatures. For the optical conductivity in the CDW regime, we also did the calculations for $T/t_0=10/13$ (which is shown in the main text) and got consistent results.
       \begin{figure}[t]
\includegraphics[width=0.99\columnwidth]{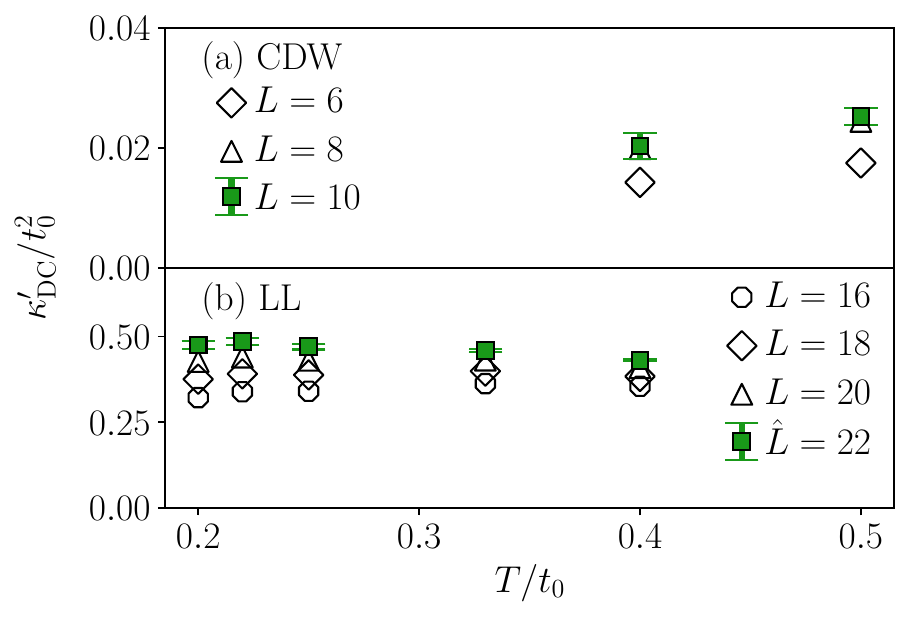}
\caption{DC thermal conductivity for the  CDW (a) and LL (b) parameters calculated with different $L$. In (a), we use $\eta=0.1/(4\pi)$ and in (b) $\eta=0.1/(4\pi)$. The errorbars for $L=10$ in (a) and $L=22$ in (b) are obtained as described in the main text of the Supplementary Material.}
\label{fig:KAP_DC}
       \end{figure}
 In the CDW parameter regime with $\eta=0.4/(4\pi)$ (shown in the main text), we find that $\Delta f\textrm{-sum}$ at maximum of order $\mathcal{O}(10^{-3})$ for both the optical and thermal conductivity at all temperatures. 
 
$\Delta f\textrm{-sum}$ is also used to determine the parameters for the Fourier transformation. We define \begin{equation} \label{eq:def_deltaf_max}
\Delta f\textrm{-sum}_{\max}=\max_T [ \Delta f\textrm{-sum} ]\, ,
\end{equation}
as the maximum difference for the aforementioned temperatures $T$.
       
In Figs.~\ref{fig:fsum_rule_comp}(a) and~\ref{fig:fsum_rule_comp}(b), we show how $\Delta f\textrm{-sum}$ varies for different values of $\eta $ and $t_{\textrm{tot}}t_0$. For the LL parameters [Fig.~\ref{fig:fsum_rule_comp}(a)], $\Delta f\textrm{-sum}$ fluctuates at small $t_{\textrm{tot}}t_0$ but saturates for later times. For the CDW parameters [Fig.~\ref{fig:fsum_rule_comp}(b)], $\Delta f\textrm{-sum}$ is roughly independent of $t_{\textrm{tot}}t_0$ (note that the times are after the initial decay). 
However, for small $\eta$, fluctuations can be seen, corresponding to the fluctuations seen in the correlation functions. Since the data are roughly stable (at the desired accuracy for our work) for $\eta=0.4/(4\pi)$, this is used in the main text.

\textit{DC conductivity error bars.} 
When computing the DC conductivity, we determine the error bars by computing $\sigma_{\textrm{DC}}^{\prime}$ ($\kappa_{\textrm{DC}}^{\prime}$) for different values of $\eta$. In Fig.~\ref{fig:kappa_dc_eta}, we show $\sigma^{\prime}_{\textrm{DC}}$ in the CDW regime (shown in the main text) for different $\eta$ and different $t_{\textrm{tot}}t_0$. $\sigma^{\prime}_{\textrm{DC}}$ seems to be roughly independent of $t_{\textrm{tot}}t_0$ (when chosen large enough) and $\eta$. The error bars are determined by fixing $\eta=0.1/(4\pi)$ [$\eta=0.4/(4\pi)$] for the LL [CDW] parameters and taking the maximum difference between $\eta=0.1/(4\pi)$ and either $\eta=0.05/(4\pi)$ or $\eta=0.15/(4\pi)$ [$\eta=0.4/(4\pi)$ and either $\eta=0.2/(4\pi)$ or $\eta=0.6/(4\pi)$].
  
Furthermore, we discard those data points where the relative uncertainty exceeds 15\%. For the data shown in the main text, the relative uncertainty is 11\% for $T/t_0=0.2$ and $<$0.7\% for the rest. Figure~\ref{fig:kappa_dc_eta}(a) demonstrates an example where the error is of the same magnitude as the quantity and our criterion does not hold, thus the data point is left out.

In Fig.~\ref{fig:KAP_DC_1}(a), we demonstrate the effect of different $\epsilon_{\textrm{bond}, \hat J^Q}$ on the DC conductivity. In the plot, some differences can be seen by changing $\epsilon_{\textrm{bond}, \hat J^Q}$ or $\eta$ (affecting the error bars). 

In Fig.~\ref{fig:KAP_DC_1}(b), we compare the DC thermal conductivity computed with $\hat J^E$ to that computed with $\hat J^E_{e-ph}$ for the LL parameters. In both cases, we first see a small increase followed by a decay when $T/t_0$ gets larger. Furthermore, going from $\hat J^E$ to  $\hat J^E_{e-ph}$  significantly decreases the DC conductivity.

$\kappa_{\textrm{DC}}^{\prime}$ can be seen in Fig.~\ref{fig:KAP_DC} for the CDW [Fig.~\ref{fig:KAP_DC}(a)] and LL [Fig.~\ref{fig:KAP_DC}(b)] parameters for different system sizes. Some finite-size effects are still visible, most prominently in the LL regime. We emphasize that our results are comparable to other results for LLs (for spin chains) in the literature \cite{huang_13,steinigeweg_15} in terms of temperatures reached (where finite-size effects also can be seen), 
 but for the more challenging case with  a significantly larger number of local degrees of freedom.
 \bibliographystyle{biblev1}
 \bibliography{references}

\begin{thebibliography}{10}
\expandafter\ifx\csname url\endcsname\relax
  \def\url#1{{\tt #1}}\fi
\expandafter\ifx\csname urlprefix\endcsname\relax\def\urlprefix{URL }\fi
\expandafter\ifx\csname bibinfo\endcsname\relax\def\bibinfo#1#2{#2}\fi
\expandafter\ifx\csname eprint\endcsname\relax\def\eprint#1{\url{#1}}\fi

\bibitem{bertini_21}
\bibinfo{author}{B.~Bertini}, \bibinfo{author}{F.~Heidrich-Meisner},
  \bibinfo{author}{C.~Karrasch}, \bibinfo{author}{T.~Prosen},
  \bibinfo{author}{R.~Steinigeweg}, and \bibinfo{author}{M.~\ifmmode
  \check{Z}\else \v{Z}\fi{}nidari\ifmmode~\check{c}\else \v{c}\fi{}},
  \bibinfo{title}{Finite-temperature transport in one-dimensional quantum
  lattice models},
  \bibinfo{journal}{\href{http://dx.doi.org/10.1103/RevModPhys.93.025003}{Rev.
  Mod. Phys.}} \href{http://dx.doi.org/10.1103/RevModPhys.93.025003}{{\bf
  \bibinfo{volume}{93}}, \bibinfo{pages}{025003}}
  (\href{http://dx.doi.org/10.1103/RevModPhys.93.025003}{\bibinfo{year}{2021}}).

\bibitem{Bulchandani2021}
\bibinfo{author}{V.~B. Bulchandani}, \bibinfo{author}{S.~Gopalakrishnan}, and
  \bibinfo{author}{E.~Ilievski}, \bibinfo{title}{Superdiffusion in spin
  chains},
  \bibinfo{journal}{\href{http://dx.doi.org/https://doi.org/10.1088/1742-5468/ac12c7}{J.
  Stat. Mech.}}
  \href{http://dx.doi.org/https://doi.org/10.1088/1742-5468/ac12c7}{\bibinfo{pages}{084001}}
  (\href{http://dx.doi.org/https://doi.org/10.1088/1742-5468/ac12c7}{\bibinfo{year}{2021}}).

\bibitem{Mahan}
\bibinfo{author}{G.~D. Mahan}, {\em \bibinfo{title}{Many-{P}article
  {P}hysics}\/} (\bibinfo{publisher}{Plenum Press}, \bibinfo{address}{New York,
  London}, \bibinfo{year}{1990}).

\bibitem{Hess2019}
\bibinfo{author}{C.~Hess}, \bibinfo{title}{Heat transport of cuprate-based
  low-dimensional quantum magnets with strong exchange coupling},
  \bibinfo{journal}{\href{http://dx.doi.org/https://doi.org/10.1016/j.physrep.2019.02.004}{Phys.
  Rep.}}
  \href{http://dx.doi.org/https://doi.org/10.1016/j.physrep.2019.02.004}{{\bf
  \bibinfo{volume}{811}}, \bibinfo{pages}{1}}
  (\href{http://dx.doi.org/https://doi.org/10.1016/j.physrep.2019.02.004}{\bibinfo{year}{2019}}).

\bibitem{perfetti_02}
\bibinfo{author}{L.~Perfetti}, \bibinfo{author}{S.~Mitrovic},
  \bibinfo{author}{G.~Margaritondo}, \bibinfo{author}{M.~Grioni},
  \bibinfo{author}{L.~Forr\'o}, \bibinfo{author}{L.~Degiorgi}, and
  \bibinfo{author}{H.~H\"ochst}, \bibinfo{title}{Mobile small polarons and the
  {Peierls} transition in the quasi-one-dimensional conductor
  {${\mathrm{K}}_{0.3}{\mathrm{MoO}}_{3}$}},
  \bibinfo{journal}{\href{http://dx.doi.org/10.1103/PhysRevB.66.075107}{Phys.
  Rev. B}} \href{http://dx.doi.org/10.1103/PhysRevB.66.075107}{{\bf
  \bibinfo{volume}{66}}, \bibinfo{pages}{075107}}
  (\href{http://dx.doi.org/10.1103/PhysRevB.66.075107}{\bibinfo{year}{2002}}).

\bibitem{li_22}
\bibinfo{author}{R.~S. Li}, \bibinfo{author}{L.~Yue}, \bibinfo{author}{Q.~Wu},
  \bibinfo{author}{S.~X. Xu}, \bibinfo{author}{Q.~M. Liu},
  \bibinfo{author}{Z.~X. Wang}, \bibinfo{author}{T.~C. Hu},
  \bibinfo{author}{X.~Y. Zhou}, \bibinfo{author}{L.~Y. Shi},
  \bibinfo{author}{S.~J. Zhang}, \bibinfo{author}{D.~Wu},
  \bibinfo{author}{T.~Dong}, and \bibinfo{author}{N.~L. Wang},
  \bibinfo{title}{Optical spectroscopy and ultrafast pump-probe study of a
  quasi-one-dimensional charge density wave in {CuTe}},
  \bibinfo{journal}{\href{http://dx.doi.org/10.1103/PhysRevB.105.115102}{Phys.
  Rev. B}} \href{http://dx.doi.org/10.1103/PhysRevB.105.115102}{{\bf
  \bibinfo{volume}{105}}, \bibinfo{pages}{115102}}
  (\href{http://dx.doi.org/10.1103/PhysRevB.105.115102}{\bibinfo{year}{2022}}).

\bibitem{Schramm_2008}
\bibinfo{author}{S.~Schramm}, \bibinfo{author}{J.~Hoffmann}, and
  \bibinfo{author}{C.~Jooss}, \bibinfo{title}{Transport and ordering of
  polarons in {CER} manganites {PrCaMnO}},
  \bibinfo{journal}{\href{http://dx.doi.org/10.1088/0953-8984/20/39/395231}{J.
  Condens. Matter Phys.}}
  \href{http://dx.doi.org/10.1088/0953-8984/20/39/395231}{{\bf
  \bibinfo{volume}{20}}, \bibinfo{pages}{395231}}
  (\href{http://dx.doi.org/10.1088/0953-8984/20/39/395231}{\bibinfo{year}{2008}}).

\bibitem{saucke_12}
\bibinfo{author}{G.~Saucke}, \bibinfo{author}{J.~Norpoth},
  \bibinfo{author}{C.~Jooss}, \bibinfo{author}{D.~Su}, and
  \bibinfo{author}{Y.~Zhu}, \bibinfo{title}{Polaron absorption for photovoltaic
  energy conversion in a manganite-titanate \textbf{pn} heterojunction},
  \bibinfo{journal}{\href{http://dx.doi.org/10.1103/PhysRevB.85.165315}{Phys.
  Rev. B}} \href{http://dx.doi.org/10.1103/PhysRevB.85.165315}{{\bf
  \bibinfo{volume}{85}}, \bibinfo{pages}{165315}}
  (\href{http://dx.doi.org/10.1103/PhysRevB.85.165315}{\bibinfo{year}{2012}}).

\bibitem{Shimshoni2003}
\bibinfo{author}{E.~Shimshoni}, \bibinfo{author}{N.~Andrei}, and
  \bibinfo{author}{A.~Rosch}, \bibinfo{title}{Thermal conductivity of spin-1/2
  chains},
  \bibinfo{journal}{\href{http://dx.doi.org/https://doi.org/10.1103/PhysRevB.68.104401}{Phys.
  Rev. B}}
  \href{http://dx.doi.org/https://doi.org/10.1103/PhysRevB.68.104401}{{\bf
  \bibinfo{volume}{68}}, \bibinfo{pages}{104401}}
  (\href{http://dx.doi.org/https://doi.org/10.1103/PhysRevB.68.104401}{\bibinfo{year}{2003}}).

\bibitem{Chernyshev2005}
\bibinfo{author}{A.~L. Chernyshev} and \bibinfo{author}{A.~V. Rozhkov},
  \bibinfo{title}{Thermal transport in antiferromagnetic spin-chain materials},
  \bibinfo{journal}{\href{http://dx.doi.org/https://doi.org/10.1103/PhysRevB.72.104423}{Phys.
  Rev. B}}
  \href{http://dx.doi.org/https://doi.org/10.1103/PhysRevB.72.104423}{{\bf
  \bibinfo{volume}{72}}, \bibinfo{pages}{104423}}
  (\href{http://dx.doi.org/https://doi.org/10.1103/PhysRevB.72.104423}{\bibinfo{year}{2005}}).

\bibitem{Boulat2007}
\bibinfo{author}{E.~Boulat}, \bibinfo{author}{P.~Mehta},
  \bibinfo{author}{N.~Andrei}, \bibinfo{author}{E.~Shimshoni}, and
  \bibinfo{author}{A.~Rosch}, \bibinfo{title}{Heat transport of clean
  spin-ladders coupled to phonons: Umklapp scattering and drag},
  \bibinfo{journal}{\href{http://dx.doi.org/https://doi.org/10.1103/PhysRevB.76.214411}{Phys.
  Rev. B}}
  \href{http://dx.doi.org/https://doi.org/10.1103/PhysRevB.76.214411}{{\bf
  \bibinfo{volume}{76}}, \bibinfo{pages}{214411}}
  (\href{http://dx.doi.org/https://doi.org/10.1103/PhysRevB.76.214411}{\bibinfo{year}{2007}}).

\bibitem{Gangadharaiah2010}
\bibinfo{author}{S.~Gangadharaiah}, \bibinfo{author}{A.~L. Chernyshev}, and
  \bibinfo{author}{W.~Brenig}, \bibinfo{title}{Thermal drag revisited:
  Boltzmann versus {Kubo}},
  \bibinfo{journal}{\href{http://dx.doi.org/10.1103/PhysRevB.82.134421}{Phys.
  Rev. B}} \href{http://dx.doi.org/10.1103/PhysRevB.82.134421}{{\bf
  \bibinfo{volume}{82}}, \bibinfo{pages}{134421}}
  (\href{http://dx.doi.org/10.1103/PhysRevB.82.134421}{\bibinfo{year}{2010}}).

\bibitem{Bartsch2013}
\bibinfo{author}{C.~Bartsch} and \bibinfo{author}{W.~Brenig},
  \bibinfo{title}{Thermal drag in spin ladders coupled to phonons},
  \bibinfo{journal}{\href{http://dx.doi.org/10.1103/PhysRevB.88.214412}{Phys.
  Rev. B}} \href{http://dx.doi.org/10.1103/PhysRevB.88.214412}{{\bf
  \bibinfo{volume}{88}}, \bibinfo{pages}{214412}}
  (\href{http://dx.doi.org/10.1103/PhysRevB.88.214412}{\bibinfo{year}{2013}}).

\bibitem{Chernyshev2015}
\bibinfo{author}{A.~L. Chernyshev} and \bibinfo{author}{W.~Brenig},
  \bibinfo{title}{Thermal conductivity in {$\text{large}\ensuremath{-}J$}
  two-dimensional antiferromagnets: Role of phonon scattering},
  \bibinfo{journal}{\href{http://dx.doi.org/10.1103/PhysRevB.92.054409}{Phys.
  Rev. B}} \href{http://dx.doi.org/10.1103/PhysRevB.92.054409}{{\bf
  \bibinfo{volume}{92}}, \bibinfo{pages}{054409}}
  (\href{http://dx.doi.org/10.1103/PhysRevB.92.054409}{\bibinfo{year}{2015}}).

\bibitem{Chernyshev2016}
\bibinfo{author}{A.~L. Chernyshev} and \bibinfo{author}{A.~V. Rozhkov},
  \bibinfo{title}{Heat transport in spin chains with weak spin-phonon
  coupling},
  \bibinfo{journal}{\href{http://dx.doi.org/10.1103/PhysRevLett.116.017204}{Phys.
  Rev. Lett.}} \href{http://dx.doi.org/10.1103/PhysRevLett.116.017204}{{\bf
  \bibinfo{volume}{116}}, \bibinfo{pages}{017204}}
  (\href{http://dx.doi.org/10.1103/PhysRevLett.116.017204}{\bibinfo{year}{2016}}).

\bibitem{kressdorf_20}
\bibinfo{author}{B.~Kressdorf}, \bibinfo{author}{T.~Meyer},
  \bibinfo{author}{A.~Belenchuk}, \bibinfo{author}{O.~Shapoval},
  \bibinfo{author}{M.~ten Brink}, \bibinfo{author}{S.~Melles},
  \bibinfo{author}{U.~Ross}, \bibinfo{author}{J.~Hoffmann},
  \bibinfo{author}{V.~Moshnyaga}, \bibinfo{author}{M.~Seibt},
  \bibinfo{author}{P.~Bl\"ochl}, and \bibinfo{author}{C.~Jooss},
  \bibinfo{title}{Room-temperature hot-polaron photovoltaics in the
  charge-ordered state of a layered perovskite oxide heterojunction},
  \bibinfo{journal}{\href{http://dx.doi.org/10.1103/PhysRevApplied.14.054006}{Phys.
  Rev. Applied}} \href{http://dx.doi.org/10.1103/PhysRevApplied.14.054006}{{\bf
  \bibinfo{volume}{14}}, \bibinfo{pages}{054006}}
  (\href{http://dx.doi.org/10.1103/PhysRevApplied.14.054006}{\bibinfo{year}{2020}}).

\bibitem{amarel_20}
\bibinfo{author}{J.~Amarel}, \bibinfo{author}{D.~Belitz}, and
  \bibinfo{author}{T.~R. Kirkpatrick}, \bibinfo{title}{Exact solution of the
  {Boltzmann} equation for low-temperature transport coefficients in metals.
  {I.} {S}cattering by phonons, antiferromagnons, and helimagnons},
  \bibinfo{journal}{\href{http://dx.doi.org/10.1103/PhysRevB.102.214306}{Phys.
  Rev. B}} \href{http://dx.doi.org/10.1103/PhysRevB.102.214306}{{\bf
  \bibinfo{volume}{102}}, \bibinfo{pages}{214306}}
  (\href{http://dx.doi.org/10.1103/PhysRevB.102.214306}{\bibinfo{year}{2020}}).

\bibitem{amarel_21}
\bibinfo{author}{J.~Amarel}, \bibinfo{author}{D.~Belitz}, and
  \bibinfo{author}{T.~R. Kirkpatrick}, \bibinfo{title}{Rigorous results for the
  electrical conductivity due to electron–phonon scattering},
  \bibinfo{journal}{\href{http://dx.doi.org/10.1063/5.0004277}{J. Math. Phys.}}
  \href{http://dx.doi.org/10.1063/5.0004277}{{\bf \bibinfo{volume}{62}},
  \bibinfo{pages}{023301}}
  (\href{http://dx.doi.org/10.1063/5.0004277}{\bibinfo{year}{2021}}).

\bibitem{fratini_01}
\bibinfo{author}{S.~Fratini}, \bibinfo{author}{F.~de~Pasquale}, and
  \bibinfo{author}{S.~Ciuchi}, \bibinfo{title}{Optical absorption from a
  nondegenerate polaron gas},
  \bibinfo{journal}{\href{http://dx.doi.org/10.1103/PhysRevB.63.153101}{Phys.
  Rev. B}} \href{http://dx.doi.org/10.1103/PhysRevB.63.153101}{{\bf
  \bibinfo{volume}{63}}, \bibinfo{pages}{153101}}
  (\href{http://dx.doi.org/10.1103/PhysRevB.63.153101}{\bibinfo{year}{2001}}).

\bibitem{fratini_03}
\bibinfo{author}{S.~Fratini} and \bibinfo{author}{S.~Ciuchi},
  \bibinfo{title}{Dynamical mean-field theory of transport of small polarons},
  \bibinfo{journal}{\href{http://dx.doi.org/10.1103/PhysRevLett.91.256403}{Phys.
  Rev. Lett.}} \href{http://dx.doi.org/10.1103/PhysRevLett.91.256403}{{\bf
  \bibinfo{volume}{91}}, \bibinfo{pages}{256403}}
  (\href{http://dx.doi.org/10.1103/PhysRevLett.91.256403}{\bibinfo{year}{2003}}).

\bibitem{fratini_06}
\bibinfo{author}{S.~Fratini} and \bibinfo{author}{S.~Ciuchi},
  \bibinfo{title}{Optical properties of small polarons from dynamical
  mean-field theory},
  \bibinfo{journal}{\href{http://dx.doi.org/10.1103/PhysRevB.74.075101}{Phys.
  Rev. B}} \href{http://dx.doi.org/10.1103/PhysRevB.74.075101}{{\bf
  \bibinfo{volume}{74}}, \bibinfo{pages}{075101}}
  (\href{http://dx.doi.org/10.1103/PhysRevB.74.075101}{\bibinfo{year}{2006}}).

\bibitem{Louis2003}
\bibinfo{author}{K.~Louis} and \bibinfo{author}{C.~Gros},
  \bibinfo{title}{Quantum {Monte Carlo} simulation for the conductance of
  one-dimensional quantum spin systems},
  \bibinfo{journal}{\href{http://dx.doi.org/10.1103/PhysRevB.68.184424}{Phys.
  Rev. B}} \href{http://dx.doi.org/10.1103/PhysRevB.68.184424}{{\bf
  \bibinfo{volume}{68}}, \bibinfo{pages}{184424}}
  (\href{http://dx.doi.org/10.1103/PhysRevB.68.184424}{\bibinfo{year}{2003}}).

\bibitem{mishcenko_03}
\bibinfo{author}{A.~S. Mishchenko}, \bibinfo{author}{N.~Nagaosa},
  \bibinfo{author}{N.~V. Prokof'ev}, \bibinfo{author}{A.~Sakamoto}, and
  \bibinfo{author}{B.~V. Svistunov}, \bibinfo{title}{Optical conductivity of
  the {Fr\"ohlich} polaron},
  \bibinfo{journal}{\href{http://dx.doi.org/10.1103/PhysRevLett.91.236401}{Phys.
  Rev. Lett.}} \href{http://dx.doi.org/10.1103/PhysRevLett.91.236401}{{\bf
  \bibinfo{volume}{91}}, \bibinfo{pages}{236401}}
  (\href{http://dx.doi.org/10.1103/PhysRevLett.91.236401}{\bibinfo{year}{2003}}).

\bibitem{mishcenko_08}
\bibinfo{author}{A.~S. Mishchenko}, \bibinfo{author}{N.~Nagaosa},
  \bibinfo{author}{Z.-X. Shen}, \bibinfo{author}{G.~De~Filippis},
  \bibinfo{author}{V.~Cataudella}, \bibinfo{author}{T.~P. Devereaux},
  \bibinfo{author}{C.~Bernhard}, \bibinfo{author}{K.~W. Kim}, and
  \bibinfo{author}{J.~Zaanen}, \bibinfo{title}{Charge dynamics of doped holes
  in high {${T}_{c}$} cuprate superconductors: A clue from optical
  conductivity},
  \bibinfo{journal}{\href{http://dx.doi.org/10.1103/PhysRevLett.100.166401}{Phys.
  Rev. Lett.}} \href{http://dx.doi.org/10.1103/PhysRevLett.100.166401}{{\bf
  \bibinfo{volume}{100}}, \bibinfo{pages}{166401}}
  (\href{http://dx.doi.org/10.1103/PhysRevLett.100.166401}{\bibinfo{year}{2008}}).

\bibitem{mischenko_15}
\bibinfo{author}{A.~S. Mishchenko}, \bibinfo{author}{N.~Nagaosa},
  \bibinfo{author}{G.~De~Filippis}, \bibinfo{author}{A.~de~Candia}, and
  \bibinfo{author}{V.~Cataudella}, \bibinfo{title}{Mobility of {Holstein}
  polaron at finite temperature: An unbiased approach},
  \bibinfo{journal}{\href{http://dx.doi.org/10.1103/PhysRevLett.114.146401}{Phys.
  Rev. Lett.}} \href{http://dx.doi.org/10.1103/PhysRevLett.114.146401}{{\bf
  \bibinfo{volume}{114}}, \bibinfo{pages}{146401}}
  (\href{http://dx.doi.org/10.1103/PhysRevLett.114.146401}{\bibinfo{year}{2015}}).

\bibitem{weber_Freericks_2021}
\bibinfo{author}{M.~Weber} and \bibinfo{author}{J.~K. Freericks},
  \bibinfo{title}{Real-time evolution of static electron-phonon models in
  time-dependent electric fields},
  \bibinfo{journal}{\href{http://dx.doi.org/10.1103/PhysRevE.105.025301}{Phys.
  Rev. E}} \href{http://dx.doi.org/10.1103/PhysRevE.105.025301}{{\bf
  \bibinfo{volume}{105}}, \bibinfo{pages}{025301}}
  (\href{http://dx.doi.org/10.1103/PhysRevE.105.025301}{\bibinfo{year}{2022}}).

\bibitem{schollwock2011density}
\bibinfo{author}{U.~Schollw\"ock}, \bibinfo{title}{The density-matrix
  renormalization group in the age of matrix product states},
  \bibinfo{journal}{\href{http://dx.doi.org/https://doi.org/10.1016/j.aop.2010.09.012}{Ann.
  Phys. (N. Y.)}}
  \href{http://dx.doi.org/https://doi.org/10.1016/j.aop.2010.09.012}{{\bf
  \bibinfo{volume}{326}}, \bibinfo{pages}{96 }}
  (\href{http://dx.doi.org/https://doi.org/10.1016/j.aop.2010.09.012}{\bibinfo{year}{2011}}).

\bibitem{zhang98}
\bibinfo{author}{C.~Zhang}, \bibinfo{author}{E.~Jeckelmann}, and
  \bibinfo{author}{S.~R. White}, \bibinfo{title}{Density matrix approach to
  local {Hilbert} space reduction},
  \bibinfo{journal}{\href{http://dx.doi.org/10.1103/PhysRevLett.80.2661}{Phys.
  Rev. Lett.}} \href{http://dx.doi.org/10.1103/PhysRevLett.80.2661}{{\bf
  \bibinfo{volume}{80}}, \bibinfo{pages}{2661}}
  (\href{http://dx.doi.org/10.1103/PhysRevLett.80.2661}{\bibinfo{year}{1998}}).

\bibitem{Guo2012}
\bibinfo{author}{C.~Guo}, \bibinfo{author}{A.~Weichselbaum},
  \bibinfo{author}{J.~von Delft}, and \bibinfo{author}{M.~Vojta},
  \bibinfo{title}{Critical and strong-coupling phases in one- and two-bath
  spin-boson models},
  \bibinfo{journal}{\href{http://dx.doi.org/10.1103/PhysRevLett.108.160401}{Phys.
  Rev. Lett.}} \href{http://dx.doi.org/10.1103/PhysRevLett.108.160401}{{\bf
  \bibinfo{volume}{108}}, \bibinfo{pages}{160401}}
  (\href{http://dx.doi.org/10.1103/PhysRevLett.108.160401}{\bibinfo{year}{2012}}).

\bibitem{brockt_dorfner_15}
\bibinfo{author}{C.~Brockt}, \bibinfo{author}{F.~Dorfner},
  \bibinfo{author}{L.~Vidmar}, \bibinfo{author}{F.~Heidrich-Meisner}, and
  \bibinfo{author}{E.~Jeckelmann}, \bibinfo{title}{Matrix-product-state method
  with a dynamical local basis optimization for bosonic systems out of
  equilibrium},
  \bibinfo{journal}{\href{http://dx.doi.org/10.1103/PhysRevB.92.241106}{Phys.
  Rev. B}} \href{http://dx.doi.org/10.1103/PhysRevB.92.241106}{{\bf
  \bibinfo{volume}{92}}, \bibinfo{pages}{241106}}
  (\href{http://dx.doi.org/10.1103/PhysRevB.92.241106}{\bibinfo{year}{2015}}).

\bibitem{stolpp2020}
\bibinfo{author}{J.~Stolpp}, \bibinfo{author}{J.~Herbrych},
  \bibinfo{author}{F.~Dorfner}, \bibinfo{author}{E.~Dagotto}, and
  \bibinfo{author}{F.~Heidrich-Meisner}, \bibinfo{title}{Charge-density-wave
  melting in the one-dimensional {Holstein} model},
  \bibinfo{journal}{\href{http://dx.doi.org/10.1103/PhysRevB.101.035134}{Phys.
  Rev. B}} \href{http://dx.doi.org/10.1103/PhysRevB.101.035134}{{\bf
  \bibinfo{volume}{101}}, \bibinfo{pages}{035134}}
  (\href{http://dx.doi.org/10.1103/PhysRevB.101.035134}{\bibinfo{year}{2020}}).

\bibitem{jansen20}
\bibinfo{author}{D.~Jansen}, \bibinfo{author}{J.~Bon\v{c}a}, and
  \bibinfo{author}{F.~Heidrich-Meisner}, \bibinfo{title}{{Finite-temperature
  density-matrix renormalization group method for electron-phonon systems:
  Thermodynamics and Holstein-polaron spectral functions}},
  \bibinfo{journal}{\href{http://dx.doi.org/10.1103/PhysRevB.102.165155}{Phys.
  Rev. B}} \href{http://dx.doi.org/10.1103/PhysRevB.102.165155}{{\bf
  \bibinfo{volume}{102}}, \bibinfo{pages}{165155}}
  (\href{http://dx.doi.org/10.1103/PhysRevB.102.165155}{\bibinfo{year}{2020}}).

\bibitem{jansen22}
\bibinfo{author}{D.~Jansen}, \bibinfo{author}{J.~Bon\ifmmode~\check{c}\else
  \v{c}\fi{}a}, and \bibinfo{author}{F.~Heidrich-Meisner},
  \bibinfo{title}{Finite-temperature optical conductivity with density-matrix
  renormalization group methods for the {Holstein} polaron and bipolaron with
  dispersive phonons},
  \bibinfo{journal}{\href{http://dx.doi.org/10.1103/PhysRevB.106.155129}{Phys.
  Rev. B}} \href{http://dx.doi.org/10.1103/PhysRevB.106.155129}{{\bf
  \bibinfo{volume}{106}}, \bibinfo{pages}{155129}}
  (\href{http://dx.doi.org/10.1103/PhysRevB.106.155129}{\bibinfo{year}{2022}}).

\bibitem{verstrate2004}
\bibinfo{author}{F.~Verstraete}, \bibinfo{author}{J.~J. Garc\'{\i}a-Ripoll},
  and \bibinfo{author}{J.~I. Cirac}, \bibinfo{title}{Matrix product density
  operators: {S}imulation of finite-temperature and dissipative systems},
  \bibinfo{journal}{\href{http://dx.doi.org/10.1103/PhysRevLett.93.207204}{Phys.
  Rev. Lett.}} \href{http://dx.doi.org/10.1103/PhysRevLett.93.207204}{{\bf
  \bibinfo{volume}{93}}, \bibinfo{pages}{207204}}
  (\href{http://dx.doi.org/10.1103/PhysRevLett.93.207204}{\bibinfo{year}{2004}}).

\bibitem{barthel2009}
\bibinfo{author}{T.~Barthel}, \bibinfo{author}{U.~Schollw\"ock}, and
  \bibinfo{author}{S.~R. White}, \bibinfo{title}{Spectral functions in
  one-dimensional quantum systems at finite temperature using the density
  matrix renormalization group},
  \bibinfo{journal}{\href{http://dx.doi.org/10.1103/PhysRevB.79.245101}{Phys.
  Rev. B}} \href{http://dx.doi.org/10.1103/PhysRevB.79.245101}{{\bf
  \bibinfo{volume}{79}}, \bibinfo{pages}{245101}}
  (\href{http://dx.doi.org/10.1103/PhysRevB.79.245101}{\bibinfo{year}{2009}}).

\bibitem{barthel2012}
\bibinfo{author}{T.~Barthel}, \bibinfo{author}{U.~Schollw\"ock}, and
  \bibinfo{author}{S.~Sachdev}, \bibinfo{title}{Scaling of the thermal spectral
  function for quantum critical bosons in one dimension},
  \href{https://arxiv.org/abs/1212.3570}{\bibinfo{howpublished}{arXiv:1212.3570}}
  (\href{http://dx.doi.org/arXiv:1212.3570}{\bibinfo{year}{2012}}).

\bibitem{barthel2013}
\bibinfo{author}{T.~Barthel}, \bibinfo{title}{Precise evaluation of thermal
  response functions by optimized density matrix renormalization group
  schemes},
  \bibinfo{journal}{\href{http://dx.doi.org/10.1088/1367-2630/15/7/073010}{New
  J. Phys.}} \href{http://dx.doi.org/10.1088/1367-2630/15/7/073010}{{\bf
  \bibinfo{volume}{15}}, \bibinfo{pages}{073010}}
  (\href{http://dx.doi.org/10.1088/1367-2630/15/7/073010}{\bibinfo{year}{2013}}).

\bibitem{karrasch2012}
\bibinfo{author}{C.~Karrasch}, \bibinfo{author}{J.~H. Bardarson}, and
  \bibinfo{author}{J.~E. Moore}, \bibinfo{title}{Finite-temperature dynamical
  density matrix renormalization group and the {Drude} weight of spin-{$1/2$}
  chains},
  \bibinfo{journal}{\href{http://dx.doi.org/10.1103/PhysRevLett.108.227206}{Phys.
  Rev. Lett.}} \href{http://dx.doi.org/10.1103/PhysRevLett.108.227206}{{\bf
  \bibinfo{volume}{108}}, \bibinfo{pages}{227206}}
  (\href{http://dx.doi.org/10.1103/PhysRevLett.108.227206}{\bibinfo{year}{2012}}).

\bibitem{karrasch2013}
\bibinfo{author}{C.~Karrasch}, \bibinfo{author}{J.~H. Bardarson}, and
  \bibinfo{author}{J.~E. Moore}, \bibinfo{title}{Reducing the numerical effort
  of finite-temperature density matrix renormalization group calculations},
  \bibinfo{journal}{\href{http://dx.doi.org/10.1088/1367-2630/15/8/083031}{New
  J. Phys.}} \href{http://dx.doi.org/10.1088/1367-2630/15/8/083031}{{\bf
  \bibinfo{volume}{15}}, \bibinfo{pages}{083031}}
  (\href{http://dx.doi.org/10.1088/1367-2630/15/8/083031}{\bibinfo{year}{2013}}).

\bibitem{kennes2016}
\bibinfo{author}{D.~Kennes} and \bibinfo{author}{C.~Karrasch},
  \bibinfo{title}{Extending the range of real time density matrix
  renormalization group simulations},
  \bibinfo{journal}{\href{http://dx.doi.org/https://doi.org/10.1016/j.cpc.2015.10.019}{Comput.
  Phys. Commun.}}
  \href{http://dx.doi.org/https://doi.org/10.1016/j.cpc.2015.10.019}{{\bf
  \bibinfo{volume}{200}}, \bibinfo{pages}{37 }}
  (\href{http://dx.doi.org/https://doi.org/10.1016/j.cpc.2015.10.019}{\bibinfo{year}{2016}}).

\bibitem{barthel_16}
\bibinfo{author}{T.~Barthel}, \bibinfo{title}{Matrix product purifications for
  canonical ensembles and quantum number distributions},
  \bibinfo{journal}{\href{http://dx.doi.org/10.1103/PhysRevB.94.115157}{Phys.
  Rev. B}} \href{http://dx.doi.org/10.1103/PhysRevB.94.115157}{{\bf
  \bibinfo{volume}{94}}, \bibinfo{pages}{115157}}
  (\href{http://dx.doi.org/10.1103/PhysRevB.94.115157}{\bibinfo{year}{2016}}).

\bibitem{paeckel_2019}
\bibinfo{author}{S.~Paeckel}, \bibinfo{author}{T.~K\"ohler},
  \bibinfo{author}{A.~Swoboda}, \bibinfo{author}{S.~R. Manmana},
  \bibinfo{author}{U.~Schollw\"ock}, and \bibinfo{author}{C.~Hubig},
  \bibinfo{title}{Time-evolution methods for matrix-product states},
  \bibinfo{journal}{\href{http://dx.doi.org/https://doi.org/10.1016/j.aop.2019.167998}{Ann.
  Phys. (N. Y.)}}
  \href{http://dx.doi.org/https://doi.org/10.1016/j.aop.2019.167998}{{\bf
  \bibinfo{volume}{411}}, \bibinfo{pages}{167998}}
  (\href{http://dx.doi.org/https://doi.org/10.1016/j.aop.2019.167998}{\bibinfo{year}{2019}}).

\bibitem{Quijad_98}
\bibinfo{author}{M.~Quijada}, \bibinfo{author}{J.~\ifmmode~\check{C}\else
  \v{C}\fi{}erne}, \bibinfo{author}{J.~R. Simpson}, \bibinfo{author}{H.~D.
  Drew}, \bibinfo{author}{K.~H. Ahn}, \bibinfo{author}{A.~J. Millis},
  \bibinfo{author}{R.~Shreekala}, \bibinfo{author}{R.~Ramesh},
  \bibinfo{author}{M.~Rajeswari}, and \bibinfo{author}{T.~Venkatesan},
  \bibinfo{title}{Optical conductivity of manganites: Crossover from
  {Jahn-Teller} small polaron to coherent transport in the ferromagnetic
  state},
  \bibinfo{journal}{\href{http://dx.doi.org/10.1103/PhysRevB.58.16093}{Phys.
  Rev. B}} \href{http://dx.doi.org/10.1103/PhysRevB.58.16093}{{\bf
  \bibinfo{volume}{58}}, \bibinfo{pages}{16093}}
  (\href{http://dx.doi.org/10.1103/PhysRevB.58.16093}{\bibinfo{year}{1998}}).

\bibitem{hartinger_06}
\bibinfo{author}{C.~Hartinger}, \bibinfo{author}{F.~Mayr},
  \bibinfo{author}{A.~Loidl}, and \bibinfo{author}{T.~Kopp},
  \bibinfo{title}{Polaronic excitations in colossal magnetoresistance manganite
  films},
  \bibinfo{journal}{\href{http://dx.doi.org/10.1103/PhysRevB.73.024408}{Phys.
  Rev. B}} \href{http://dx.doi.org/10.1103/PhysRevB.73.024408}{{\bf
  \bibinfo{volume}{73}}, \bibinfo{pages}{024408}}
  (\href{http://dx.doi.org/10.1103/PhysRevB.73.024408}{\bibinfo{year}{2006}}).

\bibitem{mildner_15}
\bibinfo{author}{S.~Mildner}, \bibinfo{author}{J.~Hoffmann},
  \bibinfo{author}{P.~E. Bl\"ochl}, \bibinfo{author}{S.~Techert}, and
  \bibinfo{author}{C.~Jooss}, \bibinfo{title}{Temperature- and doping-dependent
  optical absorption in the small-polaron system
  {$\mathrm{P}{\mathrm{r}}_{1\ensuremath{-}x}\mathrm{C}{\mathrm{a}}_{x}\mathrm{Mn}
  \mathrm{O}_{3}$}},
  \bibinfo{journal}{\href{http://dx.doi.org/10.1103/PhysRevB.92.035145}{Phys.
  Rev. B}} \href{http://dx.doi.org/10.1103/PhysRevB.92.035145}{{\bf
  \bibinfo{volume}{92}}, \bibinfo{pages}{035145}}
  (\href{http://dx.doi.org/10.1103/PhysRevB.92.035145}{\bibinfo{year}{2015}}).

\bibitem{cahill_87}
\bibinfo{author}{D.~G. Cahill} and \bibinfo{author}{R.~O. Pohl},
  \bibinfo{title}{Thermal conductivity of amorphous solids above the plateau},
  \bibinfo{journal}{\href{http://dx.doi.org/10.1103/PhysRevB.35.4067}{Phys.
  Rev. B}} \href{http://dx.doi.org/10.1103/PhysRevB.35.4067}{{\bf
  \bibinfo{volume}{35}}, \bibinfo{pages}{4067}}
  (\href{http://dx.doi.org/10.1103/PhysRevB.35.4067}{\bibinfo{year}{1987}}).

\bibitem{paddock_86}
\bibinfo{author}{C.~A. Paddock} and \bibinfo{author}{G.~L. Eesley},
  \bibinfo{title}{Transient thermoreflectance from thin metal films},
  \bibinfo{journal}{\href{http://dx.doi.org/10.1063/1.337642}{J. Appl. Phys.}}
  \href{http://dx.doi.org/10.1063/1.337642}{{\bf \bibinfo{volume}{60}},
  \bibinfo{pages}{285}}
  (\href{http://dx.doi.org/10.1063/1.337642}{\bibinfo{year}{1986}}).

\bibitem{schmidt_09}
\bibinfo{author}{A.~J. Schmidt}, \bibinfo{author}{R.~Cheaito}, and
  \bibinfo{author}{M.~Chiesa}, \bibinfo{title}{A frequency-domain
  thermoreflectance method for the characterization of thermal properties},
  \bibinfo{journal}{\href{http://dx.doi.org/10.1063/1.3212673}{Rev. Sci.
  Instrum.}} \href{http://dx.doi.org/10.1063/1.3212673}{{\bf
  \bibinfo{volume}{80}}, \bibinfo{pages}{094901}}
  (\href{http://dx.doi.org/10.1063/1.3212673}{\bibinfo{year}{2009}}).

\bibitem{Holstein1959}
\bibinfo{author}{T.~Holstein}, \bibinfo{title}{{Studies of polaron motion: Part
  I. The molecular-crystal model}},
  \bibinfo{journal}{\href{http://dx.doi.org/10.1016/0003-4916(59)90002-8}{Ann.
  Phys. (N. Y.)}} \href{http://dx.doi.org/10.1016/0003-4916(59)90002-8}{{\bf
  \bibinfo{volume}{8}}, \bibinfo{pages}{325}}
  (\href{http://dx.doi.org/10.1016/0003-4916(59)90002-8}{\bibinfo{year}{1959}}).

\bibitem{Holstein1959_2}
\bibinfo{author}{T.~Holstein}, \bibinfo{title}{{Studies of polaron motion: Part
  II. The “small” polaron}},
  \bibinfo{journal}{\href{http://dx.doi.org/https://doi.org/10.1016/0003-4916(59)90003-X}{Ann.
  Phys. (N. Y.)}}
  \href{http://dx.doi.org/https://doi.org/10.1016/0003-4916(59)90003-X}{{\bf
  \bibinfo{volume}{8}}, \bibinfo{pages}{343}}
  (\href{http://dx.doi.org/https://doi.org/10.1016/0003-4916(59)90003-X}{\bibinfo{year}{1959}}).

\bibitem{hirsch_83}
\bibinfo{author}{J.~E. Hirsch} and \bibinfo{author}{E.~Fradkin},
  \bibinfo{title}{Phase diagram of one-dimensional electron-phonon systems.
  {II.} {T}he molecular-crystal model},
  \bibinfo{journal}{\href{http://dx.doi.org/10.1103/PhysRevB.27.4302}{Phys.
  Rev. B}} \href{http://dx.doi.org/10.1103/PhysRevB.27.4302}{{\bf
  \bibinfo{volume}{27}}, \bibinfo{pages}{4302}}
  (\href{http://dx.doi.org/10.1103/PhysRevB.27.4302}{\bibinfo{year}{1983}}).

\bibitem{bursill_98}
\bibinfo{author}{R.~J. Bursill}, \bibinfo{author}{R.~H. McKenzie}, and
  \bibinfo{author}{C.~J. Hamer}, \bibinfo{title}{Phase diagram of the
  one-dimensional {Holstein} model of spinless fermions},
  \bibinfo{journal}{\href{http://dx.doi.org/10.1103/PhysRevLett.80.5607}{Phys.
  Rev. Lett.}} \href{http://dx.doi.org/10.1103/PhysRevLett.80.5607}{{\bf
  \bibinfo{volume}{80}}, \bibinfo{pages}{5607}}
  (\href{http://dx.doi.org/10.1103/PhysRevLett.80.5607}{\bibinfo{year}{1998}}).

\bibitem{creffield_05}
\bibinfo{author}{C.~E. Creffield}, \bibinfo{author}{G.~Sangiovanni}, and
  \bibinfo{author}{M.~Capone}, \bibinfo{title}{{Phonon softening and dispersion
  in the 1D Holstein model of spinless fermions}},
  \bibinfo{journal}{\href{http://dx.doi.org/10.1140/epjb/e2005-00112-9}{Eur.
  Phys. J. B}} \href{http://dx.doi.org/10.1140/epjb/e2005-00112-9}{{\bf
  \bibinfo{volume}{44}}, \bibinfo{pages}{175}}
  (\href{http://dx.doi.org/10.1140/epjb/e2005-00112-9}{\bibinfo{year}{2005}}).

\bibitem{franchini_21}
\bibinfo{author}{C.~Franchini}, \bibinfo{author}{M.~Reticcioli},
  \bibinfo{author}{M.~Setvin}, and \bibinfo{author}{U.~Diebold},
  \bibinfo{title}{Polarons in materials},
  \bibinfo{journal}{\href{http://dx.doi.org/10.1038/s41578-021-00289-w}{Nat.
  Rev. Mater.}} \href{http://dx.doi.org/10.1038/s41578-021-00289-w}{{\bf
  \bibinfo{volume}{6}}, \bibinfo{pages}{560}}
  (\href{http://dx.doi.org/10.1038/s41578-021-00289-w}{\bibinfo{year}{2021}}).

\bibitem{capone_97}
\bibinfo{author}{M.~Capone}, \bibinfo{author}{W.~Stephan}, and
  \bibinfo{author}{M.~Grilli}, \bibinfo{title}{Small-polaron formation and
  optical absorption in {Su-Schrieffer-Heeger} and {Holstein} models},
  \bibinfo{journal}{\href{http://dx.doi.org/10.1103/PhysRevB.56.4484}{Phys.
  Rev. B}} \href{http://dx.doi.org/10.1103/PhysRevB.56.4484}{{\bf
  \bibinfo{volume}{56}}, \bibinfo{pages}{4484}}
  (\href{http://dx.doi.org/10.1103/PhysRevB.56.4484}{\bibinfo{year}{1997}}).

\bibitem{zhang99}
\bibinfo{author}{C.~Zhang}, \bibinfo{author}{E.~Jeckelmann}, and
  \bibinfo{author}{S.~R. White}, \bibinfo{title}{Dynamical properties of the
  one-dimensional {Holstein} model},
  \bibinfo{journal}{\href{http://dx.doi.org/10.1103/PhysRevB.60.14092}{Phys.
  Rev. B}} \href{http://dx.doi.org/10.1103/PhysRevB.60.14092}{{\bf
  \bibinfo{volume}{60}}, \bibinfo{pages}{14092}}
  (\href{http://dx.doi.org/10.1103/PhysRevB.60.14092}{\bibinfo{year}{1999}}).

\bibitem{schubert_05}
\bibinfo{author}{G.~Schubert}, \bibinfo{author}{G.~Wellein},
  \bibinfo{author}{A.~Weisse}, \bibinfo{author}{A.~Alvermann}, and
  \bibinfo{author}{H.~Fehske}, \bibinfo{title}{Optical absorption and activated
  transport in polaronic systems},
  \bibinfo{journal}{\href{http://dx.doi.org/10.1103/PhysRevB.72.104304}{Phys.
  Rev. B}} \href{http://dx.doi.org/10.1103/PhysRevB.72.104304}{{\bf
  \bibinfo{volume}{72}}, \bibinfo{pages}{104304}}
  (\href{http://dx.doi.org/10.1103/PhysRevB.72.104304}{\bibinfo{year}{2005}}).

\bibitem{Loos2007}
\bibinfo{author}{J.~Loos}, \bibinfo{author}{M.~Hohenadler},
  \bibinfo{author}{A.~Alvermann}, and \bibinfo{author}{H.~Fehske},
  \bibinfo{title}{Optical conductivity of polaronic charge carriers},
  \bibinfo{journal}{\href{http://dx.doi.org/10.1088/0953-8984/19/23/236233}{J.
  Phys. Condens. Matter}}
  \href{http://dx.doi.org/10.1088/0953-8984/19/23/236233}{{\bf
  \bibinfo{volume}{19}}, \bibinfo{pages}{236233}}
  (\href{http://dx.doi.org/10.1088/0953-8984/19/23/236233}{\bibinfo{year}{2007}}).

\bibitem{wellein98}
\bibinfo{author}{G.~Wellein} and \bibinfo{author}{H.~Fehske},
  \bibinfo{title}{Self-trapping problem of electrons or excitons in one
  dimension},
  \bibinfo{journal}{\href{http://dx.doi.org/10.1103/PhysRevB.58.6208}{Phys.
  Rev. B}} \href{http://dx.doi.org/10.1103/PhysRevB.58.6208}{{\bf
  \bibinfo{volume}{58}}, \bibinfo{pages}{6208}}
  (\href{http://dx.doi.org/10.1103/PhysRevB.58.6208}{\bibinfo{year}{1998}}).

\bibitem{goodvin11}
\bibinfo{author}{G.~L. Goodvin}, \bibinfo{author}{A.~S. Mishchenko}, and
  \bibinfo{author}{M.~Berciu}, \bibinfo{title}{Optical conductivity of the
  {Holstein polaron}},
  \bibinfo{journal}{\href{http://dx.doi.org/10.1103/PhysRevLett.107.076403}{Phys.
  Rev. Lett.}} \href{http://dx.doi.org/10.1103/PhysRevLett.107.076403}{{\bf
  \bibinfo{volume}{107}}, \bibinfo{pages}{076403}}
  (\href{http://dx.doi.org/10.1103/PhysRevLett.107.076403}{\bibinfo{year}{2011}}).

\bibitem{weber_16}
\bibinfo{author}{M.~Weber}, \bibinfo{author}{F.~F. Assaad}, and
  \bibinfo{author}{M.~Hohenadler}, \bibinfo{title}{Thermodynamic and spectral
  properties of adiabatic {Peierls} chains},
  \bibinfo{journal}{\href{http://dx.doi.org/10.1103/PhysRevB.94.155150}{Phys.
  Rev. B}} \href{http://dx.doi.org/10.1103/PhysRevB.94.155150}{{\bf
  \bibinfo{volume}{94}}, \bibinfo{pages}{155150}}
  (\href{http://dx.doi.org/10.1103/PhysRevB.94.155150}{\bibinfo{year}{2016}}).

\bibitem{bonca_21}
\bibinfo{author}{J.~Bon\ifmmode~\check{c}\else \v{c}\fi{}a} and
  \bibinfo{author}{S.~A. Trugman}, \bibinfo{title}{Dynamic properties of a
  polaron coupled to dispersive optical phonons},
  \bibinfo{journal}{\href{http://dx.doi.org/10.1103/PhysRevB.103.054304}{Phys.
  Rev. B}} \href{http://dx.doi.org/10.1103/PhysRevB.103.054304}{{\bf
  \bibinfo{volume}{103}}, \bibinfo{pages}{054304}}
  (\href{http://dx.doi.org/10.1103/PhysRevB.103.054304}{\bibinfo{year}{2021}}).

\bibitem{fratini_21}
\bibinfo{author}{S.~Fratini} and \bibinfo{author}{S.~Ciuchi},
  \bibinfo{title}{{Displaced Drude peak and bad metal from the interaction with
  slow fluctuations.}},
  \bibinfo{journal}{\href{http://dx.doi.org/10.21468/SciPostPhys.11.2.039}{SciPost
  Phys.}} \href{http://dx.doi.org/10.21468/SciPostPhys.11.2.039}{{\bf
  \bibinfo{volume}{11}}, \bibinfo{pages}{039}}
  (\href{http://dx.doi.org/10.21468/SciPostPhys.11.2.039}{\bibinfo{year}{2021}}).

\bibitem{rezania_12}
\bibinfo{author}{H.~Rezania} and \bibinfo{author}{F.~Taherkhani},
  \bibinfo{title}{Polaron effects on the thermal conductivity of zigzag carbon
  nanotubes},
  \bibinfo{journal}{\href{http://dx.doi.org/https://doi.org/10.1016/j.ssc.2012.07.006}{Solid
  State Commun.}}
  \href{http://dx.doi.org/https://doi.org/10.1016/j.ssc.2012.07.006}{{\bf
  \bibinfo{volume}{152}}, \bibinfo{pages}{1776}}
  (\href{http://dx.doi.org/https://doi.org/10.1016/j.ssc.2012.07.006}{\bibinfo{year}{2012}}).

\bibitem{weber_assaad_18}
\bibinfo{author}{M.~Weber}, \bibinfo{author}{F.~F. Assaad}, and
  \bibinfo{author}{M.~Hohenadler}, \bibinfo{title}{Thermal and quantum lattice
  fluctuations in {Peierls} chains},
  \bibinfo{journal}{\href{http://dx.doi.org/10.1103/PhysRevB.98.235117}{Phys.
  Rev. B}} \href{http://dx.doi.org/10.1103/PhysRevB.98.235117}{{\bf
  \bibinfo{volume}{98}}, \bibinfo{pages}{235117}}
  (\href{http://dx.doi.org/10.1103/PhysRevB.98.235117}{\bibinfo{year}{2018}}).

\bibitem{supp-mat}
\eprint{Supplementary material, including a discussion of Born-Oppenheimer
  surfaces, convergence of the algo- rithm, details of the Fourier
  transformation and analysis of $f$-sum rule, extraction of the DC
  conductivity and estimation of error bars}.

\bibitem{kubo57}
\bibinfo{author}{R.~Kubo}, \bibinfo{title}{Statistical-mechanical theory of
  irreversible processes. {I.} general theory and simple applications to
  magnetic and conduction problems},
  \bibinfo{journal}{\href{http://dx.doi.org/10.1143/JPSJ.12.570}{J. Phys. Soc.
  Jpn.}} \href{http://dx.doi.org/10.1143/JPSJ.12.570}{{\bf
  \bibinfo{volume}{12}}, \bibinfo{pages}{570}}
  (\href{http://dx.doi.org/10.1143/JPSJ.12.570}{\bibinfo{year}{1957}}).

\bibitem{Luttinger_64}
\bibinfo{author}{J.~M. Luttinger}, \bibinfo{title}{Theory of thermal transport
  coefficients},
  \bibinfo{journal}{\href{http://dx.doi.org/10.1103/PhysRev.135.A1505}{Phys.
  Rev.}} \href{http://dx.doi.org/10.1103/PhysRev.135.A1505}{{\bf
  \bibinfo{volume}{135}}, \bibinfo{pages}{A1505}}
  (\href{http://dx.doi.org/10.1103/PhysRev.135.A1505}{\bibinfo{year}{1964}}).

\bibitem{pottier_2010}
\bibinfo{author}{N.~Pottier}, {\em \bibinfo{title}{Nonequilibrium Statistical
  Physics: Linear Irreversible Processes}\/} (\bibinfo{publisher}{Oxford
  University Press}, \bibinfo{address}{Oxford}, \bibinfo{year}{2010}).

\bibitem{schoenle_21}
\bibinfo{author}{C.~Sch\"onle}, \bibinfo{author}{D.~Jansen},
  \bibinfo{author}{F.~Heidrich-Meisner}, and \bibinfo{author}{L.~Vidmar},
  \bibinfo{title}{Eigenstate thermalization hypothesis through the lens of
  autocorrelation functions},
  \bibinfo{journal}{\href{http://dx.doi.org/10.1103/PhysRevB.103.235137}{Phys.
  Rev. B}} \href{http://dx.doi.org/10.1103/PhysRevB.103.235137}{{\bf
  \bibinfo{volume}{103}}, \bibinfo{pages}{235137}}
  (\href{http://dx.doi.org/10.1103/PhysRevB.103.235137}{\bibinfo{year}{2021}}).

\bibitem{huang_13}
\bibinfo{author}{Y.~Huang}, \bibinfo{author}{C.~Karrasch}, and
  \bibinfo{author}{J.~E. Moore}, \bibinfo{title}{Scaling of electrical and
  thermal conductivities in an almost integrable chain},
  \bibinfo{journal}{\href{http://dx.doi.org/10.1103/PhysRevB.88.115126}{Phys.
  Rev. B}} \href{http://dx.doi.org/10.1103/PhysRevB.88.115126}{{\bf
  \bibinfo{volume}{88}}, \bibinfo{pages}{115126}}
  (\href{http://dx.doi.org/10.1103/PhysRevB.88.115126}{\bibinfo{year}{2013}}).

\bibitem{steinigeweg_15}
\bibinfo{author}{R.~Steinigeweg}, \bibinfo{author}{J.~Gemmer}, and
  \bibinfo{author}{W.~Brenig}, \bibinfo{title}{Spin and energy currents in
  integrable and nonintegrable {spin-$\frac{1}{2}$} chains: A typicality
  approach to real-time autocorrelations},
  \bibinfo{journal}{\href{http://dx.doi.org/10.1103/PhysRevB.91.104404}{Phys.
  Rev. B}} \href{http://dx.doi.org/10.1103/PhysRevB.91.104404}{{\bf
  \bibinfo{volume}{91}}, \bibinfo{pages}{104404}}
  (\href{http://dx.doi.org/10.1103/PhysRevB.91.104404}{\bibinfo{year}{2015}}).

\bibitem{rigol_08}
\bibinfo{author}{M.~Rigol} and \bibinfo{author}{B.~S. Shastry},
  \bibinfo{title}{Drude weight in systems with open boundary conditions},
  \bibinfo{journal}{\href{http://dx.doi.org/10.1103/PhysRevB.77.161101}{Phys.
  Rev. B}} \href{http://dx.doi.org/10.1103/PhysRevB.77.161101}{{\bf
  \bibinfo{volume}{77}}, \bibinfo{pages}{161101}}
  (\href{http://dx.doi.org/10.1103/PhysRevB.77.161101}{\bibinfo{year}{2008}}).

\bibitem{haegeman_11}
\bibinfo{author}{J.~Haegeman}, \bibinfo{author}{J.~I. Cirac},
  \bibinfo{author}{T.~J. Osborne},
  \bibinfo{author}{I.~Pi\ifmmode~\check{z}\else \v{z}\fi{}orn},
  \bibinfo{author}{H.~Verschelde}, and \bibinfo{author}{F.~Verstraete},
  \bibinfo{title}{Time-dependent variational principle for quantum lattices},
  \bibinfo{journal}{\href{http://dx.doi.org/10.1103/PhysRevLett.107.070601}{Phys.
  Rev. Lett.}} \href{http://dx.doi.org/10.1103/PhysRevLett.107.070601}{{\bf
  \bibinfo{volume}{107}}, \bibinfo{pages}{070601}}
  (\href{http://dx.doi.org/10.1103/PhysRevLett.107.070601}{\bibinfo{year}{2011}}).

\bibitem{haegeman_16}
\bibinfo{author}{J.~Haegeman}, \bibinfo{author}{C.~Lubich},
  \bibinfo{author}{I.~Oseledets}, \bibinfo{author}{B.~Vandereycken}, and
  \bibinfo{author}{F.~Verstraete}, \bibinfo{title}{Unifying time evolution and
  optimization with matrix product states},
  \bibinfo{journal}{\href{http://dx.doi.org/10.1103/PhysRevB.94.165116}{Phys.
  Rev. B}} \href{http://dx.doi.org/10.1103/PhysRevB.94.165116}{{\bf
  \bibinfo{volume}{94}}, \bibinfo{pages}{165116}}
  (\href{http://dx.doi.org/10.1103/PhysRevB.94.165116}{\bibinfo{year}{2016}}).

\bibitem{itensor}
\bibinfo{author}{M.~Fishman}, \bibinfo{author}{S.~R. White}, and
  \bibinfo{author}{E.~M. Stoudenmire}, \bibinfo{title}{{The ITensor Software
  Library for Tensor Network Calculations (C++ version)}},
  \bibinfo{journal}{\href{http://dx.doi.org/10.21468/SciPostPhysCodeb.4}{SciPost
  Phys. Codebases}}
  \href{http://dx.doi.org/10.21468/SciPostPhysCodeb.4}{\bibinfo{pages}{4}}
  (\href{http://dx.doi.org/10.21468/SciPostPhysCodeb.4}{\bibinfo{year}{2022}}).

\bibitem{Reik_67}
\bibinfo{author}{H.~Reik} and \bibinfo{author}{D.~Heese},
  \bibinfo{title}{Frequency dependence of the electrical conductivity of small
  polarons for high and low temperatures},
  \bibinfo{journal}{\href{http://dx.doi.org/https://doi.org/10.1016/0022-3697(67)90089-3}{J.
  Phys. Chem. Solids}}
  \href{http://dx.doi.org/https://doi.org/10.1016/0022-3697(67)90089-3}{{\bf
  \bibinfo{volume}{28}}, \bibinfo{pages}{581}}
  (\href{http://dx.doi.org/https://doi.org/10.1016/0022-3697(67)90089-3}{\bibinfo{year}{1967}}).

\bibitem{emin_93}
\bibinfo{author}{D.~Emin}, \bibinfo{title}{Optical properties of large and
  small polarons and bipolarons},
  \bibinfo{journal}{\href{http://dx.doi.org/10.1103/PhysRevB.48.13691}{Phys.
  Rev. B}} \href{http://dx.doi.org/10.1103/PhysRevB.48.13691}{{\bf
  \bibinfo{volume}{48}}, \bibinfo{pages}{13691}}
  (\href{http://dx.doi.org/10.1103/PhysRevB.48.13691}{\bibinfo{year}{1993}}).

\bibitem{Born_27}
\bibinfo{author}{M.~Born} and \bibinfo{author}{R.~Oppenheimer},
  \bibinfo{title}{{Zur Quantentheorie der Molekeln}},
  \bibinfo{journal}{\href{http://dx.doi.org/https://doi.org/10.1002/andp.19273892002}{Ann.
  Phys. (Leipzig)}}
  \href{http://dx.doi.org/https://doi.org/10.1002/andp.19273892002}{{\bf
  \bibinfo{volume}{389}}, \bibinfo{pages}{457}}
  (\href{http://dx.doi.org/https://doi.org/10.1002/andp.19273892002}{\bibinfo{year}{1927}}).

\bibitem{Born_54}
\bibinfo{author}{M.~Born} and \bibinfo{author}{K.~Huang}, {\em
  \bibinfo{title}{Dynamical theory of crystal lattices}\/}
  (\bibinfo{publisher}{Clarendon press}, \bibinfo{address}{Oxford},
  \bibinfo{year}{1954}).

\bibitem{Bogomolov_69}
\bibinfo{author}{V.~N. Bogomolov}, \bibinfo{author}{Y.~A. Firsov},
  \bibinfo{author}{E.~K. Kudinov}, and \bibinfo{author}{D.~N. Mirlin},
  \bibinfo{title}{On the experimental observation of small polarons in rutile
  ({TiO$_2$})},
  \bibinfo{journal}{\href{http://dx.doi.org/https://doi.org/10.1002/pssb.19690350202}{physica
  status solidi (b)}}
  \href{http://dx.doi.org/https://doi.org/10.1002/pssb.19690350202}{{\bf
  \bibinfo{volume}{35}}, \bibinfo{pages}{555}}
  (\href{http://dx.doi.org/https://doi.org/10.1002/pssb.19690350202}{\bibinfo{year}{1969}}).

\bibitem{emin_69}
\bibinfo{author}{D.~Emin} and \bibinfo{author}{T.~Holstein},
  \bibinfo{title}{Studies of small-polaron motion {IV}. {Adiabatic theory of
  the Hall effect}},
  \bibinfo{journal}{\href{http://dx.doi.org/https://doi.org/10.1016/0003-4916(69)90034-7}{Ann.
  Phys. (N. Y.)}}
  \href{http://dx.doi.org/https://doi.org/10.1016/0003-4916(69)90034-7}{{\bf
  \bibinfo{volume}{53}}, \bibinfo{pages}{439}}
  (\href{http://dx.doi.org/https://doi.org/10.1016/0003-4916(69)90034-7}{\bibinfo{year}{1969}}).

\bibitem{austin_01}
\bibinfo{author}{I.~G. Austin} and \bibinfo{author}{N.~F. Mott},
  \bibinfo{title}{Polarons in crystalline and non-crystalline materials},
  \bibinfo{journal}{\href{http://dx.doi.org/10.1080/00018730110103249}{Adv.
  Phys.}} \href{http://dx.doi.org/10.1080/00018730110103249}{{\bf
  \bibinfo{volume}{50}}, \bibinfo{pages}{757}}
  (\href{http://dx.doi.org/10.1080/00018730110103249}{\bibinfo{year}{2001}}).

\bibitem{szabo_21}
\bibinfo{author}{A.~Szabó}, \bibinfo{author}{S.~A. Parameswaran}, and
  \bibinfo{author}{A.~Gopalakrishnan}, \bibinfo{title}{High-temperature
  transport and polaron speciation in the anharmonic {Holstein} model},
  \href{https://arxiv.org/abs/2110.10170}{\bibinfo{howpublished}{arXiv:2110.10170}}
  (\href{http://dx.doi.org/arXiv:2110.10170}{\bibinfo{year}{2021}}).

\bibitem{Louis2006}
\bibinfo{author}{K.~Louis}, \bibinfo{author}{P.~Prelov\ifmmode~\check{s}\else
  \v{s}\fi{}ek}, and \bibinfo{author}{X.~Zotos}, \bibinfo{title}{Thermal
  conductivity of one-dimensional spin-{$1/2$} systems coupled to phonons},
  \bibinfo{journal}{\href{http://dx.doi.org/10.1103/PhysRevB.74.235118}{Phys.
  Rev. B}} \href{http://dx.doi.org/10.1103/PhysRevB.74.235118}{{\bf
  \bibinfo{volume}{74}}, \bibinfo{pages}{235118}}
  (\href{http://dx.doi.org/10.1103/PhysRevB.74.235118}{\bibinfo{year}{2006}}).

\bibitem{Sologubenko_07}
\bibinfo{author}{A.~V. Sologubenko}, \bibinfo{author}{T.~Lorenz},
  \bibinfo{author}{H.~R. Ott}, and \bibinfo{author}{A.~Freimuth},
  \bibinfo{title}{Thermal conductivity via magnetic excitations in spin-chain
  materials},
  \bibinfo{journal}{\href{http://dx.doi.org/10.1007/s10909-007-9317-x}{J. Low
  Temp. Phys.}} \href{http://dx.doi.org/10.1007/s10909-007-9317-x}{{\bf
  \bibinfo{volume}{147}}, \bibinfo{pages}{387}}
  (\href{http://dx.doi.org/10.1007/s10909-007-9317-x}{\bibinfo{year}{2007}}).

\bibitem{sentef13}
\bibinfo{author}{M.~Sentef}, \bibinfo{author}{A.~F. Kemper},
  \bibinfo{author}{B.~Moritz}, \bibinfo{author}{J.~K. Freericks},
  \bibinfo{author}{Z.-X. Shen}, and \bibinfo{author}{T.~P. Devereaux},
  \bibinfo{title}{Examining electron-boson coupling using time-resolved
  spectroscopy},
  \bibinfo{journal}{\href{http://dx.doi.org/10.1103/PhysRevX.3.041033}{Phys.
  Rev. X}} \href{http://dx.doi.org/10.1103/PhysRevX.3.041033}{{\bf
  \bibinfo{volume}{3}}, \bibinfo{pages}{041033}}
  (\href{http://dx.doi.org/10.1103/PhysRevX.3.041033}{\bibinfo{year}{2013}}).

\bibitem{kemper_17}
\bibinfo{author}{A.~F. Kemper}, \bibinfo{author}{M.~A. Sentef},
  \bibinfo{author}{B.~Moritz}, \bibinfo{author}{T.~P. Devereaux}, and
  \bibinfo{author}{J.~K. Freericks}, \bibinfo{title}{Review of the theoretical
  description of time-resolved angle-resolved photoemission spectroscopy in
  electron-phonon mediated superconductors},
  \bibinfo{journal}{\href{http://dx.doi.org/10.1002/andp.201600235}{Ann. Phys.
  (Berl.)}} \href{http://dx.doi.org/10.1002/andp.201600235}{{\bf
  \bibinfo{volume}{529}}, \bibinfo{pages}{1600235}}
  (\href{http://dx.doi.org/10.1002/andp.201600235}{\bibinfo{year}{2017}}).

\bibitem{paeckel_fauseweh_19}
\bibinfo{author}{S.~Paeckel}, \bibinfo{author}{B.~Fauseweh},
  \bibinfo{author}{A.~Osterkorn}, \bibinfo{author}{T.~K\"ohler},
  \bibinfo{author}{D.~Manske}, and \bibinfo{author}{S.~R. Manmana},
  \bibinfo{title}{Detecting superconductivity out of equilibrium},
  \bibinfo{journal}{\href{http://dx.doi.org/10.1103/PhysRevB.101.180507}{Phys.
  Rev. B}} \href{http://dx.doi.org/10.1103/PhysRevB.101.180507}{{\bf
  \bibinfo{volume}{101}}, \bibinfo{pages}{180507}}
  (\href{http://dx.doi.org/10.1103/PhysRevB.101.180507}{\bibinfo{year}{2020}}).

\bibitem{rincon_21}
\bibinfo{author}{J.~Rinc\'on} and \bibinfo{author}{A.~E. Feiguin},
  \bibinfo{title}{Nonequilibrium optical response of a one-dimensional {Mott}
  insulator},
  \bibinfo{journal}{\href{http://dx.doi.org/10.1103/PhysRevB.104.085122}{Phys.
  Rev. B}} \href{http://dx.doi.org/10.1103/PhysRevB.104.085122}{{\bf
  \bibinfo{volume}{104}}, \bibinfo{pages}{085122}}
  (\href{http://dx.doi.org/10.1103/PhysRevB.104.085122}{\bibinfo{year}{2021}}).

\bibitem{ejima_22}
\bibinfo{author}{S.~Ejima}, \bibinfo{author}{F.~Lange}, and
  \bibinfo{author}{H.~Fehske}, \bibinfo{title}{Nonequilibrium dynamics in
  pumped {Mott} insulators},
  \bibinfo{journal}{\href{http://dx.doi.org/10.1103/PhysRevResearch.4.L012012}{Phys.
  Rev. Research}}
  \href{http://dx.doi.org/10.1103/PhysRevResearch.4.L012012}{{\bf
  \bibinfo{volume}{4}}, \bibinfo{pages}{L012012}}
  (\href{http://dx.doi.org/10.1103/PhysRevResearch.4.L012012}{\bibinfo{year}{2022}}).

\bibitem{jeckelmann98}
\bibinfo{author}{E.~Jeckelmann} and \bibinfo{author}{S.~R. White},
  \bibinfo{title}{Density-matrix renormalization-group study of the polaron
  problem in the {Holstein} model},
  \bibinfo{journal}{\href{http://dx.doi.org/10.1103/PhysRevB.57.6376}{Phys.
  Rev. B}} \href{http://dx.doi.org/10.1103/PhysRevB.57.6376}{{\bf
  \bibinfo{volume}{57}}, \bibinfo{pages}{6376}}
  (\href{http://dx.doi.org/10.1103/PhysRevB.57.6376}{\bibinfo{year}{1998}}).

\bibitem{schroeder_16}
\bibinfo{author}{F.~A. Y.~N. Schr\"oder} and \bibinfo{author}{A.~W. Chin},
  \bibinfo{title}{Simulating open quantum dynamics with time-dependent
  variational matrix product states: {T}owards microscopic correlation of
  environment dynamics and reduced system evolution},
  \bibinfo{journal}{\href{http://dx.doi.org/10.1103/PhysRevB.93.075105}{Phys.
  Rev. B}} \href{http://dx.doi.org/10.1103/PhysRevB.93.075105}{{\bf
  \bibinfo{volume}{93}}, \bibinfo{pages}{075105}}
  (\href{http://dx.doi.org/10.1103/PhysRevB.93.075105}{\bibinfo{year}{2016}}).

\bibitem{bischoff_17}
\bibinfo{author}{J.-M. Bischoff} and \bibinfo{author}{E.~Jeckelmann},
  \bibinfo{title}{Density-matrix renormalization group method for the
  conductance of one-dimensional correlated systems using the {Kubo} formula},
  \bibinfo{journal}{\href{http://dx.doi.org/10.1103/PhysRevB.96.195111}{Phys.
  Rev. B}} \href{http://dx.doi.org/10.1103/PhysRevB.96.195111}{{\bf
  \bibinfo{volume}{96}}, \bibinfo{pages}{195111}}
  (\href{http://dx.doi.org/10.1103/PhysRevB.96.195111}{\bibinfo{year}{2017}}).

\bibitem{yang_20}
\bibinfo{author}{M.~Yang} and \bibinfo{author}{S.~R. White},
  \bibinfo{title}{Time-dependent variational principle with ancillary {Krylov}
  subspace},
  \bibinfo{journal}{\href{http://dx.doi.org/10.1103/PhysRevB.102.094315}{Phys.
  Rev. B}} \href{http://dx.doi.org/10.1103/PhysRevB.102.094315}{{\bf
  \bibinfo{volume}{102}}, \bibinfo{pages}{094315}}
  (\href{http://dx.doi.org/10.1103/PhysRevB.102.094315}{\bibinfo{year}{2020}}).

\bibitem{koehler20}
\bibinfo{author}{T.~Köhler}, \bibinfo{author}{J.~Stolpp}, and
  \bibinfo{author}{S.~Paeckel}, \bibinfo{title}{Efficient and flexible approach
  to simulate low-dimensional quantum lattice models with large local {Hilbert}
  spaces},
  \bibinfo{journal}{\href{http://dx.doi.org/10.21468/SciPostPhys.10.3.058}{SciPost
  Phys.}} \href{http://dx.doi.org/10.21468/SciPostPhys.10.3.058}{{\bf
  \bibinfo{volume}{10}}, \bibinfo{pages}{58}}
  (\href{http://dx.doi.org/10.21468/SciPostPhys.10.3.058}{\bibinfo{year}{2021}}).

\bibitem{Xu_21}
\bibinfo{author}{Y.~Xu}, \bibinfo{author}{Z.~Xie}, \bibinfo{author}{X.~Xie},
  \bibinfo{author}{U.~Schollw\"ock}, and \bibinfo{author}{H.~Ma},
  \bibinfo{title}{Stochastic adaptive single-site time-dependent variational
  principle},
  \bibinfo{journal}{\href{http://dx.doi.org/10.1021/jacsau.1c00474}{JACS Au}}
  \href{http://dx.doi.org/10.1021/jacsau.1c00474}{{\bf \bibinfo{volume}{2}},
  \bibinfo{pages}{335}}
  (\href{http://dx.doi.org/10.1021/jacsau.1c00474}{\bibinfo{year}{2022}}).

\bibitem{mardazad_21}
\bibinfo{author}{S.~Mardazad}, \bibinfo{author}{Y.~Xu},
  \bibinfo{author}{X.~Yang}, \bibinfo{author}{M.~Grundner},
  \bibinfo{author}{U.~Schollw\"ock}, \bibinfo{author}{H.~Ma}, and
  \bibinfo{author}{S.~Paeckel}, \bibinfo{title}{Quantum dynamics simulation of
  intramolecular singlet fission in covalently linked tetracene dimer},
  \bibinfo{journal}{\href{http://dx.doi.org/10.1063/5.0068292}{J. Chem. Phys.}}
  \href{http://dx.doi.org/10.1063/5.0068292}{{\bf \bibinfo{volume}{155}},
  \bibinfo{pages}{194101}}
  (\href{http://dx.doi.org/10.1063/5.0068292}{\bibinfo{year}{2021}}).

\bibitem{moroder_22}
\bibinfo{author}{M.~Moroder}, \bibinfo{author}{M.~Grundner},
  \bibinfo{author}{F.~Damanet}, \bibinfo{author}{U.~Schollw\"ock},
  \bibinfo{author}{S.~Mardazad}, \bibinfo{author}{S.~Flannigan},
  \bibinfo{author}{T.~K\"ohler}, and \bibinfo{author}{S.~Paeckel},
  \bibinfo{title}{Stable bipolarons in open quantum systems},
  \bibinfo{journal}{\href{http://dx.doi.org/10.1103/PhysRevB.107.214310}{Phys.
  Rev. B}} \href{http://dx.doi.org/10.1103/PhysRevB.107.214310}{{\bf
  \bibinfo{volume}{107}}, \bibinfo{pages}{214310}}
  (\href{http://dx.doi.org/10.1103/PhysRevB.107.214310}{\bibinfo{year}{2023}}).

\end{thebibliography}

\end{document}